\documentclass[12pt]{article}
\usepackage[margin=1in]{geometry}
\usepackage{amsmath}
\usepackage{graphicx}
\usepackage{float}
\usepackage{cite}
\usepackage{inputenc}
\usepackage{amssymb}
\usepackage{authblk}
\usepackage[normalem]{ulem}
\usepackage{xcolor}
\usepackage{mathtools}
\usepackage{cancel}
\usepackage{caption}
\usepackage{subcaption}
\usepackage{amssymb}
\usepackage[font=scriptsize]{caption}
\usepackage[toc]{appendix}
\usepackage{color}

\title{Exploring holographic Composite Higgs models}
\author[1,2]{Djuna Croon}
\author[1]{Barry M. Dillon}
\author[1]{Stephan J. Huber}
\author[1]{Veronica Sanz}
\affil[1]{\textit{\small{Department of Physics and Astronomy, University of Sussex, BN1 9QH Brighton, UK}}}
\affil[2]{\textit{\small{Perimeter Institute for Theoretical Physics, Waterloo, ON, Canada}}}

\begin{document}
\maketitle
\begin{abstract}
Simple Composite Higgs models predict new vector-like fermions not too far from the electroweak scale, yet LHC limits are now sensitive to the TeV scale. Motivated by this tension, we explore the holographic dual of the minimal model, MCHM$_5$, to understand how far naive 4D predictions are from their 5D duals. Interestingly, we find that the usual hierarchy among the vector-like quarks is not generic, hence ameliorating the tuning issue. We find that lowering the UV cutoff in the 5D picture allows for heavier top partners, while keeping the mass of the Higgs boson at its observed value. In the 4D dual this corresponds to increasing the number of ``colours" $N$.  This is essentially a `Little Randall-Sundrum Model', which are known to reduce some flavour and electroweak constraints.  Furthermore, in anticipation of the ongoing efforts at the LHC to put bounds on the top Yukawa, we demonstrate that deviations from the SM can be suppressed or enhanced with respect to what is expected from mere symmetry arguments in 4D.
We conclude that the 5D holographic realisation of the MCHM$_5$ with a small UV cutoff is not in tension with the current experimental data. 
\end{abstract} 

\section{Introduction}
The discovery of the Higgs boson at LHC \cite{Aad:2012tfa,Chatrchyan:2012xdj}, which so-far appears to be very Standard Model-like, and the non observation of new physics raises the question of the origin of the electroweak scale.   Among some well motivated explanations of this scale, such as supersymmetry or extra dimensions, is Higgs compositeness.  In this framework the Higgs field is composed of some particles interacting via a strongly interacting gauge theory which confines at the TeV scale.  A ``little hierarchy" between the electroweak scale and the new physics scale can arise naturally if the Higgs bound state is a pseudo-Goldstone boson of this new strongly interacting sector.  

Despite difficulties in extracting predictions from strongly coupled gauge theories, several methods have been developed.  The most basic of these makes use of large $N$ approximations in $SU(N)$ gauge theories, and of the global symmetry structure in the low energy effective theory \cite{DeCurtis:2011yx,Matsedonskyi:2012ym,Marzocca:2012zn,Pomarol:2012qf,Giudice:2007fh,Anastasiou:2009rv,Panico:2011pw,Azatov:2011qy}.  These methods, although useful, can be rather limited since precise calculations of form factors are impracticable.  This means that one is unable to extract precise values of physical quantities, such as the mass spectrum of particles and their couplings.  For these reasons it is difficult to constrain the models using experimental data.

It is possible to make progress beyond this using computational tools such as lattice simulations, and while determining baryon states is still challenging, some studies in non-minimal Composite Higgs models have been done regarding the structure of the meson states~\cite{Lewis:2011zb,Hietanen:2013fya,Hietanen:2014xca}.  In this paper we adopt another popular method, namely holography, which has been proven useful to describe another strongly coupled theory, QCD at low energies~\cite{Son:2003et,Erlich:2005qh,DaRold:2005zs,Hirn:2005nr,Hirn:2005vk} as well as a way to develop new, non-QCD like, models of Technicolor~\cite{Csaki:2003zu,Cacciapaglia:2004jz,Hirn:2006nt,Hirn:2006wg}. In the context of Composite Higgses, the pioneer papers of Contino et al.~\cite{Agashe:2004rs,Contino:2010rs}, followed an intense exploration of the Higgs as a holographic pseudo-Goldstone boson in warped extra-dimensions, see e.g.~\cite{Contino:2003ve}.  Holography  is a method based on the conjectured duality between strongly interacting gauge theories in 4D and weakly coupled gravitational theories on a 5D AdS space.  Since the dual theory is weakly coupled, we are able to extract precise predictions for the form factors and all masses and couplings in the model. Here the word {\it precision} comes from the determination of 4D observables in terms of the 5D model parameters after dimensional reduction, yet the relation with the target strongly coupled 4D theory is still a conjecture and hence bound to an inherent uncertainty. 

The physics of 5D AdS spaces \cite{Randall:1999ee,ArkaniHamed:1999dc,Huber:2000ie,Huber:2000fh} was studied independently of its  application to composite Higgs models, and many of the results and constraints are the same in both cases.  The most important of these are the constraints imposed by the electroweak precision observables.  In the absence of additional symmetries, large corrections to the $T$ parameter imply a lower bound on the spin-1 resonances of $\sim 8$ TeV \cite{Dillon:2014zea,Cabrer:2011fb}.  Some ways of improving this bound are to account for incalculable contributions to the operators in the 5D bulk \cite{Dillon:2014zea}, consider modifications to the AdS geometry \cite{Cabrer:2011fb}, employ models with more than one extra dimension \cite{Archer:2010hh}, or introduce large brane kinetic terms for gauge fields \cite{Carmona:2014iwa}.  The most natural way to protect larger contributions to the $T$ parameter however, is to extend the gauge sector in the bulk to include a custodial symmetry \cite{Agashe:2003zs,Carena:2007ua}.  This mechanism is employed in most realistic composite Higgs models, and allows for spin-1 resonances with masses as low as about $2.5$ TeV. 

The space of composite Higgs models is parametrised by the global symmetry structure of the low energy effective theory, and the embedding of the quarks and leptons into this global symmetry.  A large literature exists on the simplest composite Higgs models.  We will focus on what is known as the Minimal Composite Higgs Model (MCHM) \cite{Agashe:2004rs,Contino:2010rs} with the quarks and leptons embedded in fundamental representations of the global symmetry ($\text{MCHM}_5$) \cite{Contino:2006qr,Archer:2014qga}.  This model has a global $SO(5)$ symmetry broken to $SO(4)$ at the TeV scale, thus employing the custodial protection of the $T$ parameter.  A detailed discussion of this model is reserved for section 2.  For further details on the model-building approaches in Composite Higgs models see \cite{Agashe:2005dk,Panico:2015jxa,vonGersdorff:2015fta,Low:2015nqa,Barbieri:2015lqa,Cheng:2013qwa,Barnard:2014tla,Cheng:2014dwa,Ferretti:2014qta,DeCurtis:2014oza,Carmona:2015haa}.  

Using the 4D approach it has been shown that it is possible to reproduce the correct top mass, Higgs vev and Higgs mass quite naturally.  However it is found that this usually requires light top partners.  Calculations using holography confirm this and provide more precise predictions \cite{Contino:2006qr,Archer:2014qga}.  Typically top partners below about $700$ GeV are required, and this is already in tension with bounds on vector-like quarks at the LHC~\cite{Khachatryan:2015oba,Aad:2015kqa} which, for single channel final states, already reach 900 GeV.  For specific information on top partner phenomenology we refer the reader to \cite{DeSimone:2012fs,Backovic:2015bca,Buckley:2015nca,Serra:2015xfa,Dawson:2013uqa,Grojean:2013qca,Backovic:2014ega,Drueke:2014pla,Reuter:2014iya,Matsedonskyi:2014mna,Gripaios:2014pqa,Chen:2014xwa,Banfi:2013yoa} and for general LHC phenomenology of the MCHM to \cite{Carena:2014ria,Niehoff:2015iaa,Barnard:2015ryq,Cacciapaglia:2015eqa,Thamm:2015zwa,Kanemura:2014kga,Vignaroli:2014bpa,Redi:2013eaa,Agashe:2009bb,Carmona:2012jk,Redi:2012ha,Contino:2011np}.  

There have been some attempts to alleviate the need for the light top partners.  Using 4D calculations it has been shown that this can be done by employing more complex embeddings for the third generation quarks \cite{Pomarol:2012qf,Panico:2012uw}. Instead using the holographic approach it has been shown that by embedding the leptons in larger representations, their contributions to the Higgs potential can help alleviate the need for light top partners \cite{Carmona:2014iwa}.  Also using the holographic realisations (although with a flat background), authors in \cite{Pappadopulo:2013vca} use larger embeddings for the third generation to reduce the fine-tuning in the Higgs potential and allow for heavier top-partners.  More recently, models of composite Higgs with more than one breaking scale have been studied, and it was found that this also allows for heavier top partners \cite{Sanz:2015sua}.

The correspondence between the 4D and 5D models can be described in terms of a dictionary from which one can relate the 4D and 5D parameters.  One entry in this relates the number of ``colours" in the strongly coupled 4D gauge theory to the UV cutoff in the 5D AdS theory.  In this work we investigate how the top partner spectrum changes as we vary this parameter.  The effects of lowering this UV cutoff has been studied previously in 5D AdS scenarios in which the Higgs is not a pseudo-Goldstone boson, these models are referred to as ``Little Randall-Sundrum" models \cite{Bauer:2008xb,Davoudiasl:2008hx}.  It has been shown that these models reduce bounds on some flavour and electroweak observables.  In models of gauge-Higgs unification, lowering the UV cutoff allows for lower values of $v/f_{\pi}$ while keeping the Kaluza-Klein (KK) scale constant.  In the dual theory this is related to an increase in the number of colours ``$N$".  In doing this we find that, while keeping the KK scale and the Higgs and top quark masses at the observed values, we can increase the mass of the lightest top resonance.  This is easily understood in the KK picture, where lowering the UV scale modifies the couplings of the KK modes.

Having constructed a MCHM without light top partners, we investigate deviations in the top Yukawa coupling, motivated by the ongoing experimental effort at LHC to put bounds on deviations from the SM prediction.  In composite Higgs models the top Yukawa is generally suppressed compared to the SM. If this effect is too large, it could lead to a potential conflict with current or future data. We study the top Yukawa coupling in the 5D realisation and find that in some regions of parameter space the deviations to the SM can be suppressed relative to what is expected from pure (4D) symmetry arguments. This will be very relevant once the experimental precision on the top Yukawa increases.

Overall, we find that our 5D holographic realisation of the $\text{MCHM}_5$ with a smaller UV cutoff is not in tension with current experimental data (both on the top partner spectrum and the top Yukawa coupling). The mechanisms we study that allow for heavier top partners and suppressed Yukawa deviations are very general, and in particular do not rely on any specific coset or fermion embedding.  Therefore, we expect that these results will generalise to non-minimal versions of composite Higgs, and it will be interesting and fruitful to study this in detail in the future.

\section{Overview of the $\text{MCHM}_5$}
\label{MCHMoverview}
Composite Higgs models posit a new strong sector with a global symmetry ($SO(5)\times U(1)_X$ in the MCHM) which spontaneously breaks to its subgroup ($SO(4)\times U(1)_X$), below its confinement scale. The resulting four Goldstone bosons, transforming in the fundamental representation of $SO(4)$ (or equivalently as a bi-doublet of $SU(2)_L\times SU(2)_R$), are identified with the Higgs doublet. A tree-level potential for the Goldstone bosons is forbidden by shift-symmetry, but a potential is generated radiatively if we introduce a further explicit breaking. This is done by gauging the subgroup $SU(2)_L\times U(1)_Y$ of $SO(4)\times U(1)_X$ and by choice of the fermion interaction structure. 

The new strong sector adds heavy bound states, with masses around the breaking scale, to the Standard Model field content. Mixing between the new states and the SM results in modified couplings; constraints can be placed on these modifications normalised to the SM prediction. Most stringently, electroweak precision tests put bounds the gauge boson self-energy parametrised by the oblique parameters. In the MCHM the $T$-parameter is protected from large corrections due to the custodial symmetry.
However, as we will see, the $S$-parameter bounds form an important constraint on the scale of new spin-1 resonances. 

The spectrum of the spin-1 states is fixed by the symmetry breaking pattern, but there is some freedom for the new spin-$\frac{1}{2}$ states. One has to choose how to embed the standard model quarks and leptons into the $SO(5)\times U(1)_X$ symmetry, and how to introduce an explicit breaking.  As the third generation of the SM couples most strongly to the Higgs, we will focus on that for our present work, as is customary.  

Embedding the standard model $SU(2)_L$ doublets in bi-doublets of $SU(2)_L\times SU(2)_R$ protects the $Zb\bar{b}$ from large corrections.  A simple way to do this is to embed each standard model quark generation into two fundamentals of $SO(5)\times U(1)_X$,
\begin{equation} \label{eq:quarkmultiplets}
\xi_{q1} = \begin{bmatrix}
\psi^{\prime}_{q1,(L,R)} \\
\psi_{q1,(L,R)} \\
\eta_{q1,(L,R)}  \end{bmatrix}_{\frac{2}{3}},
\xi_{q2} = \begin{bmatrix}
\psi_{q2,(L,R)}\\
\psi^{\prime}_{q2,(L,R)} \\
\eta_{q2,(L,R)}\end{bmatrix}_{-\frac{1}{3}}
\end{equation}
where the $\psi$ fields transform as a bi-doublet of $SU(2)_L\times SU(2)_R$ and the $\eta$ fields transform as singlets. The elements of each multiplet have left and right-handed components such that the new fermionic states couple to the gauge fields in a vector-like way. 
The Standard Model left-handed doublets are identified with one linear combination of $\psi_{q1,L}$ and $\psi_{q2,L}$, while the other linear combination is made massive.  The right-handed fields are identified with $\eta_{q1,R}$ and $\eta_{q2,R}$.  The charge under $U(1)_X$ is assigned such that the fields carry the correct hypercharge, $Y=T^3_R+X$. 

As the SM is the low energy limit of the theory, the non-SM fields are assumed to have masses of the order of the breaking scale. The SM fields have heavy spin-$\frac{1}{2}$ partners with the same quantum numbers. The spurious fields give rise to  
additional exotic multiplets with charges $Y=\frac{7}{6}$ and $Y=-\frac{5}{3}$.

Fermionic contributions to the Higgs potential are introduced via linear $SO(5)$ violating couplings to heavy composite fermionic degrees of freedom.  This mechanism is known as partial compositeness.  The same couplings are also responsible for generating the masses and Yukawa couplings of the SM fields.

It has long been known that the 4D model described here has a dual in 5D gauge-Higgs unification. The strong coupling in the 4D action makes it impossible to compute the form factors perturbatively, but the weak coupling in the dual allows one to calculate them explicitly. 
5D methods therefore provide very useful analytical tools for studying strongly coupled 4D gauge theories. In the next section we will describe in more detail a 5D model leading to the low energy physics as the $\text{MCHM}_5$ described in this section.

\section{A holographic model}
\label{holographicmodel}
In this section we follow closely the calculational procedure of \cite{Archer:2014qga}.  We will consider a 5D AdS bulk space bounded by two 3-branes, 
\begin{equation}\label{eq:lineelement}
ds^2=\frac{R^2}{r^2}(\eta^{\mu\nu}dx_{\mu}dx_{\nu}-dr^2),
\end{equation} 
where $r$ is a conformal co-ordinate related to the fifth spatial co-ordinate, $y$, by $r=\frac{1}{k}e^{ky}$, in which $k$ is the curvature of the 5D space.  The branes are located at $r = R = 1/k$ (the UV) and $r=R^{\prime}=\frac{1}{M_{KK}}\sim\mathcal{O}($TeV$^{-1})$ (the IR).  The position of the IR brane is related to the scale of spontaneous symmetry breaking in 4D.  

In the MCHM dual, the 5D bulk has a local $SO(5)\times U(1)_X$ gauge symmetry.  To describe the third quark generation, we require four fermion multiplets living in the 5D bulk, transforming as fundamentals of $SO(5)$.  Two of these correspond to $\xi_{q1}$ and $\xi_{q2}$, with $U(1)_X$ charges $\frac{2}{3}$ and $-\frac{1}{3}$.  And the other two, $\xi_{u}$ and $\xi_{d}$, correspond to the composite states required by the partial compositeness mechanism.   The boundary conditions of the 5D fields are assigned as follows,
\begin{equation}\label{eq:quarkmultiplets5D}\nonumber
A^{a}(++), \hspace{4mm}A^{\hat{a}}(+-)
\end{equation}
\begin{equation} 
\xi_{q1} = \begin{bmatrix}
\psi^{\prime}_{q1} (-+)\\
\psi_{q1} (++) \\
\eta_{q1} (--) \end{bmatrix}_{\frac{2}{3}}, \hspace{2mm}
\xi_{q2} = \begin{bmatrix}
\psi_{q2} (++)\\
\psi^{\prime}_{q2} (-+) \\
\eta_{q2} (--) \end{bmatrix}_{-\frac{1}{3}}
\end{equation}
\begin{equation}\nonumber
\xi_{u} = \begin{bmatrix}
\psi^{\prime}_{u} (+-)\\
\psi_{u} (+-) \\
\eta_{u} (-+) \end{bmatrix}_{\frac{2}{3}},\hspace{2mm}
\xi_{d} = \begin{bmatrix}
\psi_{d} (+-)\\
\psi^{\prime}_{d} (+-) \\
\eta_{d} (-+) \end{bmatrix}_{-\frac{1}{3}}
\end{equation}
where $A^{a}$ and $A^{\hat{a}}$ are the $SO(4)\times U(1)_X$ and broken generators, respectively.  Here the $+$ ($-$) represents a Neumann (Dirichlet) boundary condition, and the order of these is to be understood as (UV, IR).  For the gauge fields we denote the boundary condition on the $(\mu,\nu)$ component, while the $A^5$ components will have the opposite boundary conditions.  For the fermion fields we denote the boundary condition on the left-handed mode, while the right-handed modes will also have the opposite boundary conditions.  It follows that fermion fields with $(++)$ will have a massless left-handed component, while those with $(--)$ have a massless right-handed component, and fields with $(+-)$ or $(-+)$ have no massless components. For the gauge fields, components with $(++)$ boundary conditions will have a massless $A^{\mu}$ mode, while components with $(--)$ will have a massless $A^5$ scalar mode, and again the components with $(+-)$ or $(-+)$ will not have any massless component.  

The $SO(5)\times U(1)_X$ symmetry on the UV brane should be broken to the SM electroweak group in such a way that $Y=T^3_R+X$.  In addition to this, the linear combination $(\psi_{q1,L}-\psi_{q2,L})$ should be given a mass on the UV brane so that only $(\psi_{q1,L}+\psi_{q2,L})$ has a massless component.  We then identify the SM left-handed doublet as $\psi_q=(\psi_{q1,L}+\psi_{q2,L})$.  Taking all of this into account, the low energy theory now looks very much like the SM before electroweak symmetry breaking.

\subsection{ The 5D gauge sector}

In the 5D models the composite Higgs can be identified with the zero mode of the fifth component of the 5D gauge fields, i.e. $A_5^0$.  The only $A_5$ fields which have a massless zero mode are those with $(--)$ boundary conditions.  From eq.~\ref{eq:quarkmultiplets5D} it can be seen that these precisely correspond to the $SO(5)/SO(4)$ generators, as expected.  With the description of the model given so far, the dynamics of the 5D gauge sector is fixed.  The only free parameters being the scale $M_{KK}$ and the ratio of UV/IR scales, $\ln(\Omega)$.  In principle we could also include brane kinetic terms, but these are expected to be radiatively induced and we assume their effects to be negligible.  

From the 5D model one can derive expressions for observables in the 4D composite Higgs model.  Firstly, the decay constant of the Goldstone fields is found to be,
\begin{equation}\label{eq:fpi}
f_{\pi}^2=\frac{4M_{KK}^2}{g^2\ln{\Omega}}
\end{equation}
where $g$ is the EW gauge coupling $\sim0.65$ and $\Omega=\frac{R^\prime}{R}$ is the ratio of scales in the model. This expression can be generalized to non-AdS warped metrics as in Refs.~\cite{Hirn:2007bb,Randall:2002tg,Hirn:2007we}.

 Since the Higgs doublet is a Goldstone field, its couplings are of the form $\sin(h/f_{\pi})$.  Once electroweak symmetry is broken, obtaining the correct W and Z boson masses requires that
\begin{equation}\label{eq:sh}
\sin^2\left(\frac{h}{f_{\pi}}\right)\equiv s_h^2=\frac{v^2}{f_{\pi}^2}=\frac{m_W^2}{M_{KK}^2}\ln(\Omega).
\end{equation}
Thus deviations in the Higgs boson couplings can depend as much on the UV scale as they do the IR.  The holographic picture relates the $1/N$ (large number of ``colours") expansion in a 4D strongly coupled gauge theory to an expansion in a small 5D gauge coupling $g_5$ in AdS space.  From this picture the following correspondence arises,
\begin{equation}\label{eq:Ng5}
\frac{1}{N}=\frac{g_5^2k}{16\pi^2},
\end{equation}
where $g_5^2k=g^2\ln(\Omega)$. This allows us to think of the ratio of scales in the 5D theory as dual to the number of colours in the 4D picture: larger $N$ implies a smaller $\ln(\Omega)$, which is also related to the cutoff of the theory~\cite{Randall:2002tg}.   Note that there is no reason to keep $\Omega\sim 10^{15}$, as is done in some warped extra dimensional models to solve the Planck-electroweak hierarchy problem. We should remember that the 5D NDA condition for calculability requires that $\frac{g^2\ln(\Omega)}{24\pi^2}<<1$, but even with $\ln(\Omega)=40$ this is $\sim 0.07$.  Lastly, an important bound on these models comes in the form of the electroweak $S$-parameter, which can be expressed as,
\begin{equation}\label{eq:sparam}
S\simeq\frac{3}{8}\frac{N}{\pi}s_h^2=\frac{6\pi}{g^2\ln{\Omega}}s_h^2=\frac{3\pi v^2}{2M_{KK}^2}.
\end{equation}
The constraints for $S$ are correlated with the $T$ parameter: for an exact custodial symmetry, $T=0$, the bound is $S<0.02$ (implying $M_{KK}>3.8$ TeV), allowing for maximal contribution to $T$ it relaxes to $S<0.3$ (which is saturated for $M_{KK}\simeq 1$ TeV). In this latter case the lightest gauge KK modes are approximately at $\frac{3\pi}{4}M_{KK}\sim2.35$ TeV.  Note that as long as some hierarchy exists between the IR and UV scales, the $S$-parameter only depends on the IR scale.

\subsection{The 5D quark sector}

In the quark sector, to familiarise ourselves with the parameters of the model, it is instructive to look at the 5D action for the fields,
\begin{equation}\label{eq:fermionaction}
S_{\Phi}=\int d^4x\int_R^{R^{\prime}}dr\sqrt{|g|}\sum_{i=q,u}\left(\frac{1}{2}\left(\bar{\xi}_i\gamma^MD_M\xi_i-D_M\bar{\xi}_i\gamma^M\xi_i\right)-c_{i}k\bar{\xi}_i\xi_i\right),
\end{equation}
\begin{equation} \nonumber
+\hspace{2mm} \int d^4x\left(m_u\bar{\xi}^b_q\xi^b_u+M_u\bar{\xi}^s_q\xi^s_u+h.c.\right)_{r=R^{\prime}}
\end{equation}
where $E_a^M$ is the f\"{u}nfbein, $E_a^M\gamma^a=\gamma^{M}$, $\gamma^a=(\gamma^{\mu},i\gamma^5)$ are the gamma matrices in flat space, and $\omega_M$ is the spin connection.  The $b$ and $s$ superscripts in the brane mixing terms denote the bi-doublet and singlet components of the fermion multiplets. The IR brane masses control the breaking of $SO(5)$: for $m_u\neq 1/M_u$ it is broken explicitly, preserving the $SO(4)$ subgroup. 
 
Varying the 5D mass parameters ($c_q \text{ and } c_u$) determines the degree of compositeness of the fermionic operators. The field $\xi_q$ has a left-handed zero mode and hence becomes more composite as $c_q$ moves in a negative direction.  Whereas $\xi_u$ has a right-handed zero mode so becomes more composite as $c_u$ moves in a positive direction.  For $c_q=-c_u$ the fields have the same degree of compositeness. In the Kaluza-Klein picture these mass parameters control the localisation of any massless zero modes in the spectrum: a greater composite component corresponds to more IR localisation.

We have defined $\psi_q=(\psi_{q1,L}+\psi_{q2,L})$, but when calculating the Higgs potential the state with the most composite mixing will contribute the most.  Thus when calculating the top quark contribution to the Higgs potential we will assume $\psi_{q1,L}$ to be most composite and take $\psi_{q}\simeq \psi_{q1,L}$.

\subsection{The effective action}
Once the model is defined, one can write down the most general effective Lagrangian compatible with the symmetry structure.  In the case of the $\text{MCHM}_5$ this is,
\begin{equation} \nonumber
\mathcal{L}_{\text{eff}}=-\frac{P_t^{\mu\mu}}{2}\left[\frac{2}{g_5^2}W_{\mu}^+\left( \Pi_0+\frac{s_h^2}{2}\Pi_1\right)W_{\nu}^{-}  +  A_{\mu}\left(\frac{1}{g_5^2}\Pi_0 + \frac{c_w^2-s_x^2}{g_{5,X}^2}\Pi_0^{X}\right)A_{\nu} \right.
\end{equation}
\begin{equation} \nonumber
\left. + Z_{\mu}\left(\frac{c_w^2+s_x^2s_w^2}{g_5^2} \Pi_0+\frac{c_x^2s_w^2}{g_{5,X}^2} \Pi_0^X+\frac{s_h^2}{2c_w^2g_5^2}\Pi_1\right)Z_{\nu} \right]+\bar{q}_L\left(\Pi^q_0+\frac{s_h^2}{2}\Pi^q_1H^cH^{c\dagger}\right)\cancel{p}q_L
\end{equation}
\begin{equation}
+\bar{u}_R\left(\Pi^u_0+\frac{s_h^2}{2}\Pi^u_1\right)\cancel{p}u_R+\frac{s_hc_h}{\sqrt{2}}M^u_1\bar{q}_LH^cu_R + h.c.
\end{equation}
The form factors $\Pi_0^X$ are associated with the $U(1)_X$ gauge field, and the mixing angles $s_x$ and $c_x$ arise via the breaking to the SM subgroup on the UV brane.  For more details on the how this effective action is derived we refer the reader to \cite{Archer:2014qga}.  In a 4D approach one can only estimate the momentum dependence of these form factors based on sum-rules and Large-$N$ gauge theory results.  But in the 5D holographic approach they can be explicitly calculated, the results of these calculations are presented in the appendix.  It is expected that the form factors will contain poles corresponding to massive composite resonances at $p^2\sim M^2_{KK}$.  In the 5D approach these are simply the Kaluza-Klein states one obtains from the 5D gauge and fermion fields.  While the masses of the spin-1 resonances are solely determined by $M_{KK}$, the masses of the spin-$\frac{1}{2}$ resonances depend also on $c_q$, $c_u$, $m_u$ and $M_u$.  Before EWSB, when $s_h=0$, they can be expressed in terms of the above form factors as,
\begin{equation}
\begin{gathered}
m_{1/6}=\text{zeros}\{\cancel{p}\Pi^q_0\} \\
m_{2/3}=\text{zeros}\{\cancel{p}\Pi^u_0\} \\ 
m_{7/6}=\text{poles}\{\cancel{p}(\Pi^u_0+\Pi^u_1)\}.
\end{gathered}\label{eq:mT}
\end{equation}
After EWSB the $(1/6)$ and $(2/3)$ states mix resulting in a tower of top partners with $(2/3)$ charge and masses determined by the zeros of,
\begin{equation}
\left[p^2\left(\Pi^q_0+\frac{s_h^2}{2}\Pi^q_1\right)\left(\Pi^u_0+\frac{s_h^2}{2}\Pi^u_1\right)-\frac{s_h^2c_h^2}{2}(M^u_1)^2\right] \label{eq:mEWSB}.
\end{equation}
There will also be a tower of states with hypercharge $(5/3)$ and mass equal to $m_{7/6}$.  It is generally found that when one or both of the multiplets has a large composite mixing, there will generally be relatively light fermionic states in the model.  This large compositeness also generally implies a large gap in the masses of the lightest $(1/6)$, $(2/3)$ and $(7/6)$ top partners.  Thus, by varying the 5D mass parameters, we can significantly alter the spectrum of top partners we expect to observe.  Summarising, from the 5D description of the model we have six parameters, 
\begin{equation}
M_{KK}  \hspace{5mm}  \ln{\Omega}  \hspace{5mm}  c_q  \hspace{5mm}  c_u  \hspace{5mm}  m_u  \hspace{5mm}  M_u .
\end{equation}  
We can use three observables to fit to: $v$, $m_h$ and $m_t$, leaving us with three free parameters. Here we will demonstrate the freedom that these parameters give in the top sector.  In particular, there are three aspects we wish to study,
\begin{itemize}
\item How the 5D parameters are related to the top partner masses;
\item How the top partner masses are related to $s_h$, and;
\item How much 5D contributions alter the top Yukawa deviation expected from 4D composite Higgs models. 
\end{itemize}

\section{Higgs potential and EWSB}
From the effective action for the gauge fields and the top quark it is a simple exercise to write down the Coleman-Weinberg expression for the one-loop Higgs potential. After a Wick rotation we arrive at the following field-dependent potential, 
\begin{equation}\label{eq:vh}
V(h)=\int\frac{d^4p_E}{(2\pi)^4}\left( \frac{3}{2}\log\left[1+\frac{3}{2}\frac{\Pi_1}{\Pi_0}\right] - 6 \log\left[ \left( 1+\frac{s_h^2}{2}\frac{\Pi^q_1}{\Pi^q_0} \right)  \left( 1+\frac{s_h^2}{2}\frac{\Pi^u_1}{\Pi^u_0} \right) + \frac{s_h^2c_h^2}{2}\frac{(M^u_1)^2}{p_E^2\Pi^q_0\Pi^u_0}  \right]   \right)
\end{equation}
where we have neglected the effects of the $U(1)_X$ field. Expanding these logs, it is found that the potential has the following form,
\begin{equation}\label{eq:vha}
V(h)\simeq(\alpha_G+\alpha_F)s_h^2-\beta_F s_h^2c_h^2
\end{equation}
where the $F$ and $G$ subscripts refer to gauge and fermion contributions.  Notice that without the fermion contribution one cannot achieve EWSB at all.  Minimising this we find that the Higgs potential has a non-trivial ground state when $\beta_F>0$ and $\beta_F>|\alpha_F+\alpha_G|$, situated at
\begin{equation}\label{eq:vev}
s_h^2=\frac{1}{2}-\frac{\alpha_G+\alpha_F}{2\beta_F}.
\end{equation}
Taking the second derivative of $V(h)$ we find,
\begin{equation}\label{eq:mh}
m_H^2=\frac{8\beta_F}{f_{\pi}^2}s_h^2c_h^2.
\end{equation}
After EWSB it is found that the mass of the top is given by,
\begin{equation}\label{eq:mt}
m_t^2\simeq\frac{s_h^2c_h^2}{2}\frac{(M^u_1)^2}{(\Pi^q_0+\frac{s_h^2}{2}\Pi^q_1)(\Pi^u_0+\frac{s_h^2}{2}\Pi^u_1)}\Bigg|_{p^2=(174 \text{GeV})^2}.
\end{equation}
Since the top quark gives by far the most dominant contribution to the potential, we should expect a lot of correlation between the top partner spectrum and the Higgs mass.  Approximating the form factors by their limiting expressions for vanishing momentum, we can write this in terms of the 5D parameters as 
\begin{equation}\label{eq:mt5D} m_t^2\simeq\frac{M_u v \sqrt{(\tilde{c}_q-2) \tilde{c}_q (\tilde{c}_u-2) \tilde{c}_u} \sqrt{1-\frac{v^2}{f_{\pi}^2}} (1-m_u M_u)}{f_{\pi} L_1 \sqrt{-(\tilde{c}_q-2) M_u^2+\frac{\tilde{c}_u v^2 \left(m_u^2 M_u^2-1\right)}{f_{\pi}^2}+\tilde{c}_u} \sqrt{M_u^2 \left(\tilde{c}_q m_u^2 \left(2-\frac{v^2}{f_{\pi}^2}\right)-2 \tilde{c}_u+4\right)+\frac{\tilde{c}_q v^2}{f_{\pi}^2}}}, \end{equation} 
where we have defined
\begin{equation}\label{eq:5DmassSimp} c_u =  \frac{\tilde{c}_u-1}{2}\,\,\,\text{    and      }\,\,\,\,c_q = \frac{1-\tilde{c}_q}{2}, \end{equation}
such that $0 \leq \tilde{c}_{q}$, and $\tilde{c}_{u}  \leq 2$, and the profiles are flat ($c_{q,u} = \pm 1/2$) for $\tilde{c}_{q,u} = 0$.

\subsection{Yukawa couplings in the holographic $\text{MCHM}_5$}
From the discussion above it is seen that the Yukawa coupling of the top quark in $\text{MCHM}_5$ deviates from its Standard Model value. 
Following the definition of the effective Yukawa coupling by \cite{Carena:2014ria},
\begin{equation}\label{eq:yt}
y_{\psi}^{(0)}\simeq\frac{dm_{\psi}^{(0)}}{dv},
\end{equation}
we will be interested in the quantity
\begin{equation}\label{eq:kappat}
 \kappa_t = \frac{y_{t}^{(0)}}{y_{t,SM}^{(0)}} = \frac{y_{t}^{(0)}v}{m_{t}^{(0)}}.
\end{equation}
The current LHC ATLAS bounds are $\kappa_t = 0.94\pm 0.21$ at $2 \sigma$~\cite{Aad:2015gba}. This bound is expected to be strengthened to the ten percent level after the current run.  

From \eqref{eq:mt} we may calculate $\kappa_t$ in terms of the 5D form factors. To quartic order in $s_h = v/f_{\pi}$, we have
\begin{equation}\label{eq:kappat5D}
\kappa_t= 1-\frac{s_h^2}{c_h^2}-s_h^2 \left(\frac{\Pi_1^q}{2 \Pi_0^q}+\frac{\Pi_1^u}{2 \Pi_0^u}\right)+s_h^4 \left(\frac{(\Pi_1^q)^2}{4( \Pi_0^q)^2}+\frac{(\Pi_1^u)^2}{4 (\Pi_0^u)^2}\right)+O\left(s_h^5\right).
\end{equation}
As by definition, the Standard Model result ($\kappa_t=1$) is recovered in the limit $s_h\rightarrow0$. Also, as we have noted above, if the IR brane masses are related as $M_u=-1/m_u$, the fermion form factors vanish ($\Pi_1^q=\Pi_1^u=0$). In this case the BSM Yukawa corrections are universal and equal to $-s_h^2/c_h^2$ (to all orders in $s_h$).  From \eqref{eq:mt5D}, in terms of the fermion profiles we have,
\begin{align}
&\frac{y_{\psi}^{(0)}v}{m_{\psi}^{(0)}}= 
1 -\frac{s_h^2}{c_h^2} - s_h^2 \left(\frac{\left(m_u^2 M_u^2-1\right) \left(M_u^2 \left(\tilde{c}_q \left((2-\tilde{c}_q)-2 \tilde{c}_u m_u^2\right)-2 (2-\tilde{c}_u) \tilde{c}_u\right)+\tilde{c}_q \tilde{c}_u\right)}{2 M_u^2 \left(-\tilde{c}_q m_u^2-(2-\tilde{c}_u)\right) \left((2-\tilde{c}_q) M_u^2+\tilde{c}_u\right)}
\right) + O\left(s_h^4 \right).
\end{align}
In section 6 we will study how these additional contributions proportional to $\left(m_u^2 M_u^2-1\right)$ can play a role in alleviating tensions with bounds from the LHC.

\section{Phenomenology of the holographic Composite Higgs}
In this section we explore some phenomenological consequences of the Holographic Higgs, including the mass scale of the top partners, deviations to the top Yukawa coupling and possible future measurements of the Higgs in association with a hard object (vector boson, jet) as a probe for the Higgs-top-antitop form factor.

\subsection{Does a light Higgs imply light top partners?}
Taking the values of $s_h$ and $c_h$ at the minimum of $V(h)$, we can re-write the Higgs mass term from eq.~\ref{eq:mh} as,
\begin{equation} \label{eq:mh2}
m_H^2=\frac{2}{f_{\pi}^2}\frac{\beta^2-\alpha^2}{\beta}.
\end{equation}
The $\alpha$ and $\beta$ terms are of dimension four and we can expect them to be $\sim M_{KK}^4$.  Thus to obtain a light Higgs we require a degree of cancellation among the terms in the Higgs potential.  A similar cancellation is also required to obtain a light vacuum expectation value.  Due to the required cancellation among these terms, the precise value of $s_h$ alone is only a crude estimate of the fine-tuning of the model.

It has been shown that if $M_{KK}\sim 1$ TeV, and $f_{\pi}\sim 500$ GeV, one requires the $\xi_u$ multiplet to have a large composite mixing in order to get the correct degree of cancellation in $\alpha$ and $\beta$, and thus obtain the correct values of $m_H$, $m_{t,pole}$ and $v$ \cite{Contino:2006qr}.  This implies that light top partners are expected in model, with a large gap among the different charged states.  Light top partners are also a feature expected from the 4D models, where $\alpha$ and $\beta$ are estimated in terms of top partner masses.  Currently, this prediction is in tension with observations at the LHC. 

The obvious way to avoid these constraints is to push up $M_{KK}$, but in doing one severely increases the fine-tuning of the model and it becomes ``un-natural".  There have been several attempts at alleviating the need for light top partners without increasing the fine-tuning, in both the purely 4D and the holographic picture. An example of the former is \cite{Pomarol:2012qf,Panico:2012uw}, in which the authors show that by embedding the third generation in different representations of SO(5), the structure of the Higgs mass term can be altered.  For particular cases a light Higgs could be obtained with top partners $\sim 1$ TeV in this way.  The authors point out that to achieve a light Higgs with moderate fine-tuning, it is preferred to have $m_T/f_{\pi}\sim 1$, where $m_T$ is the scale of the top partner masses. To highlight an example of a holographic approach, in \cite{Carmona:2014iwa} the realisation of the model includes leptonic contributions to the Higgs potential, which allow the authors to show that a light Higgs can be achieved while having top partners $\sim 1$ TeV, with only moderate fine-tuning.

In this paper we wish to investigate an alternative method of reducing the need for light top partners in the holographic realisation of the model.  Moving the top zero mode wave functions away from the IR brane increases the mass of the top partners, but simultaneously results in an increase in the Higgs mass.  However, by lowering the UV scale (i.e. lowering $\ln\left(\Omega\right)$) we increase $f_{\pi}$ and suppress the Higgs mass.  Using this mechanism we can push the top zero mode wave functions further from the IR, pushing up the top partner masses, while keeping the Higgs mass at the observed value.  In the 4D dual, lowering the UV scale should correspond to an increase in the number of colours ``N" of the strongly coupled gauge theory \cite{Davoudiasl:2008hx,Contino:2006qr}.

\begin{figure}[H]
\centering
\begin{minipage}{.48\textwidth}
  \includegraphics[width=\textwidth]{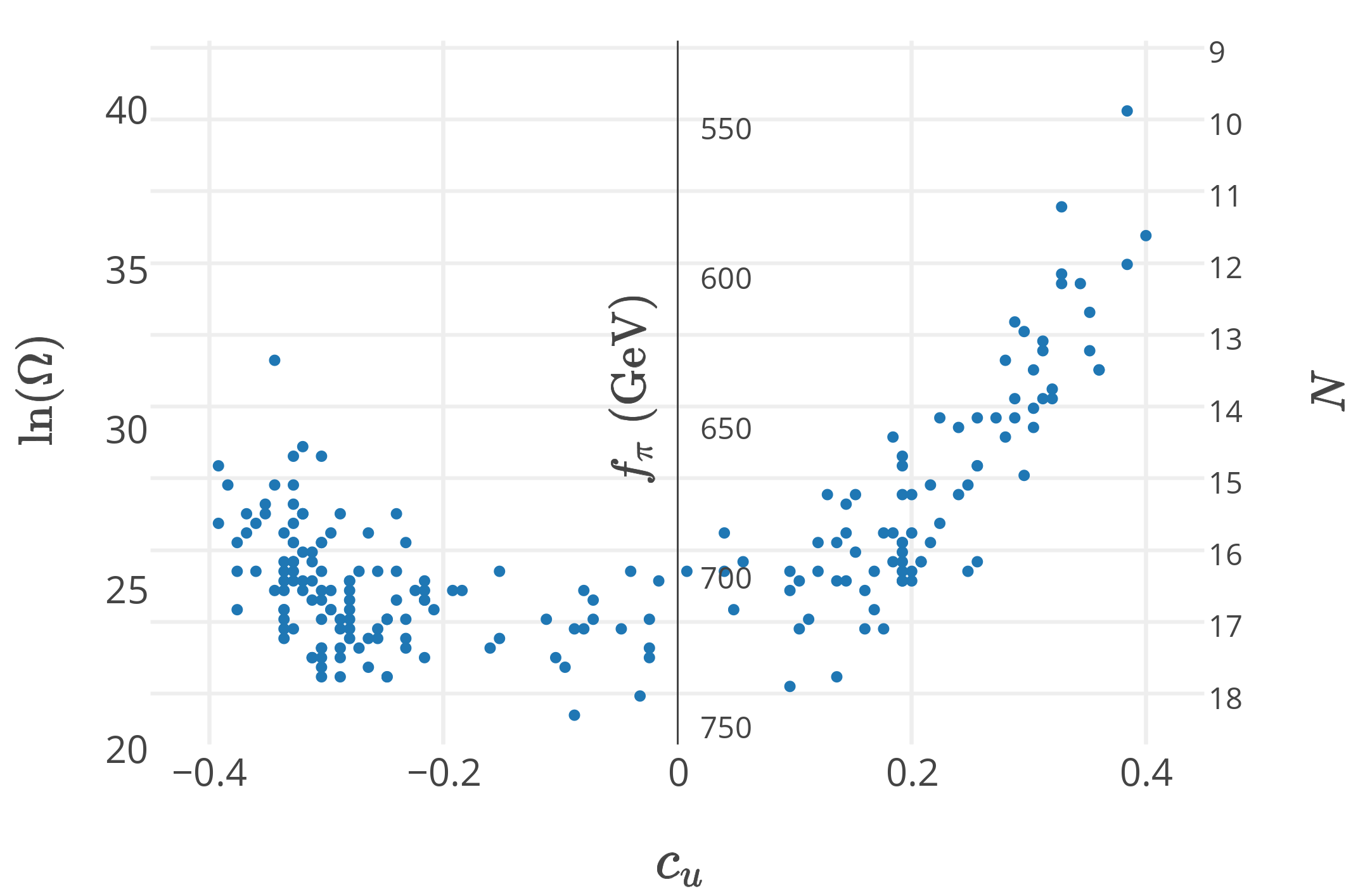}
  \caption{Correlation between $c_u$ and $\ln(\Omega)$ when $|m_u|<1.4$.  \label{fig:cuOmegaA}}
\end{minipage}%
\hfill
\begin{minipage}{.48\textwidth}
  \includegraphics[width=\textwidth]{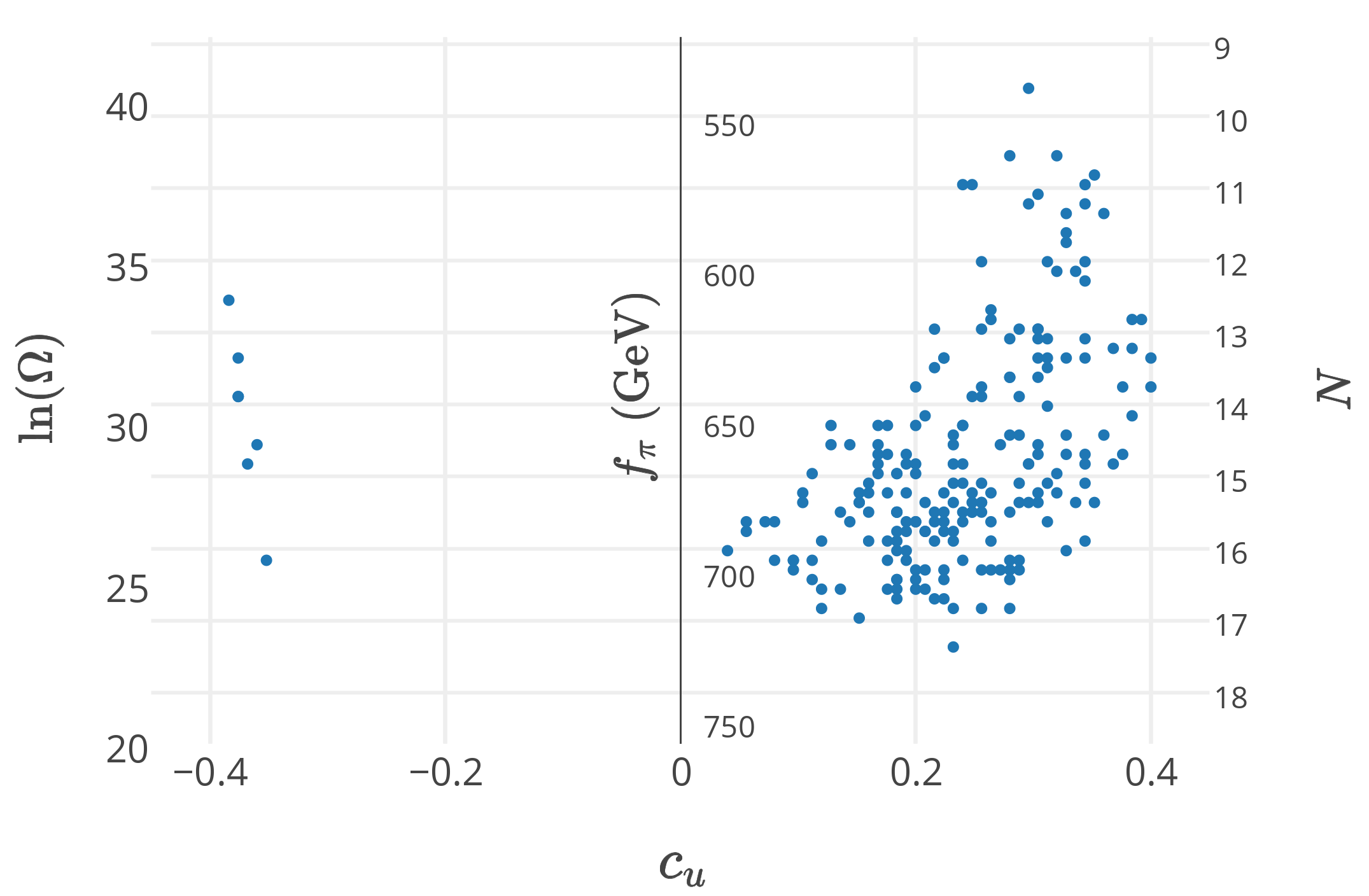}
  \caption{Correlation between $c_u$ and $\ln(\Omega)$ when $|m_u|>1.4$. \label{fig:cuOmegaB}}
\end{minipage}
\end{figure}

To illustrate this idea we perform a scan in which we fix $M_{KK}=1.1$ TeV and vary the other parameters in the ranges $0.2<c_q<0.4$, $-0.4<c_u<0.4$, $-2<m_uM_u<-0.5$ and $20<\ln(\Omega) <50$.  For $c_q=0.5$ ($c_u=-0.5$) the 5D profile of the left-handed (right-handed) zero mode will be flat.  So the choices of fermion localisations ensure that the composite mixing for $q_L$ is small, whereas the mixing of the $t_R$ state is allowed to be large or small.  We find two distinct cases in the results, $|m_u|<1.4$ and $|m_u|>1.4$.  In figures~\ref{fig:cuOmegaA} and~\ref{fig:cuOmegaB} below we show how $c_u$ and $\ln(\Omega)$ are correlated after we fix $m_{t,pole}$, $m_H$, and $v$ to their observed values.
From these plots it is clear that for a large value of $\ln(\Omega)$ ($\gtrsim 35$), a light Higgs requires the spurious multiplet to have large positive values of $c_u$.  However by allowing for smaller values of $\ln(\Omega)$ we can have significantly different values for this $c_u$ parameter.  The effects of this on the top partner spectrum are shown below in figures~\ref{fig:cumTA} and~\ref{fig:cumTB}. 
\begin{figure}[H]
\centering
\begin{minipage}{.48\textwidth}
  \includegraphics[width=\textwidth]{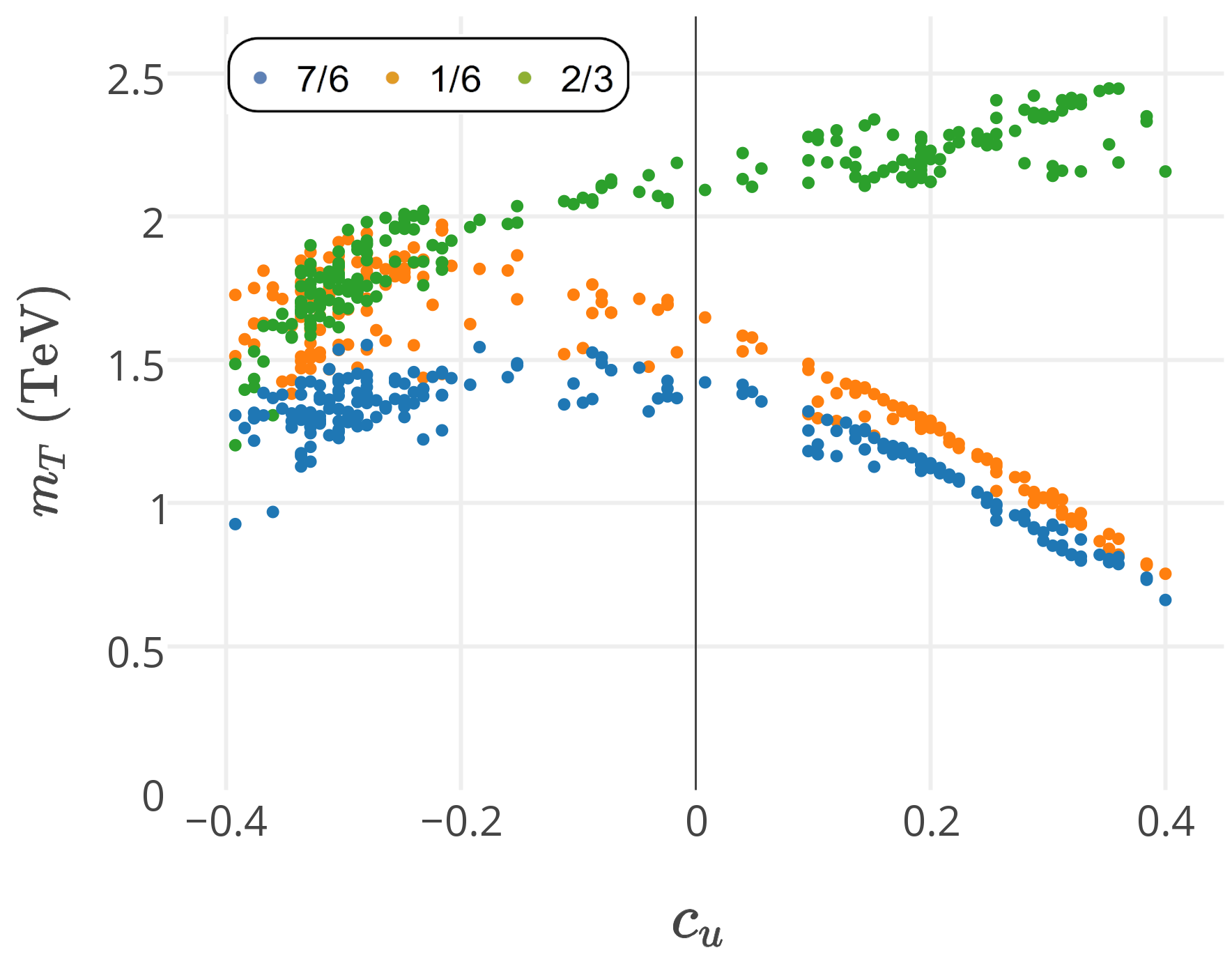}
  \caption{Correlation between $c_u$ and the top partner masses when $|m_u|<1.4$. Here the green points correspond to the top partner with hypercharge (2/3), the orange with (1/6), and the blue with (7/6). \label{fig:cumTA}}
\end{minipage}%
\hfill
\begin{minipage}{.48\textwidth}
  \includegraphics[width=\textwidth]{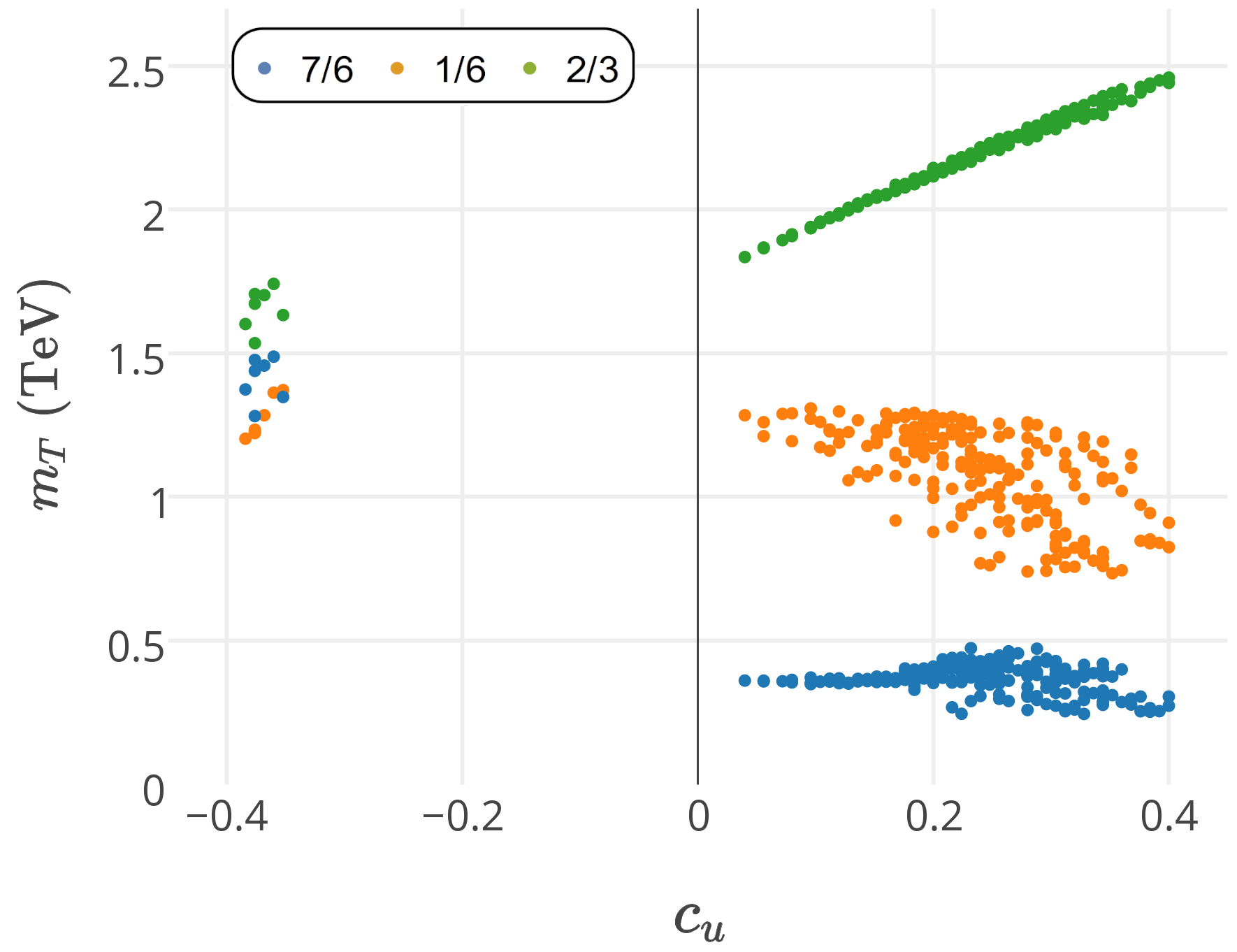}
  \caption{Correlation between $c_u$ and the top partner masses when $|m_u|>1.4$. As in the left panel, the different coloured points correspond to top partners with different hypercharge. \label{fig:cumTB}}
\end{minipage}
\end{figure}
If we were to fix $\ln(\Omega)$ to be $> 35$, we would be forced to have $c_u\gtrsim 0.3$.  This results in a distinct top partner spectrum in which the left-handed top partner and exotic top partners are $\lesssim 1$ TeV while the right-handed top partner is $\sim 2$ TeV.  However, by lowering the value of $\ln(\Omega)$ we can move $c_u$ to regions with less composite mixing in which the top partner spectrum is remarkably different.  We can easily have scenarios where all the top partners have masses $\gtrsim 1$ TeV, and where the mass gap among the different charged states is very small.  Note that, in the 4D picture, having $\ln(\Omega)\sim 37$ means having the number of colours at $\sim 10$.  Lowering $\ln(\Omega)$ to $\sim 25$ means that $N\sim 15$.  In the case of figure~\ref{fig:cumTA}, we can say that the mass gap between the top partners is strongly related to their degree of compositeness.

\begin{figure}[H]
\centering
\begin{minipage}{.48\textwidth}
  \includegraphics[width=\textwidth]{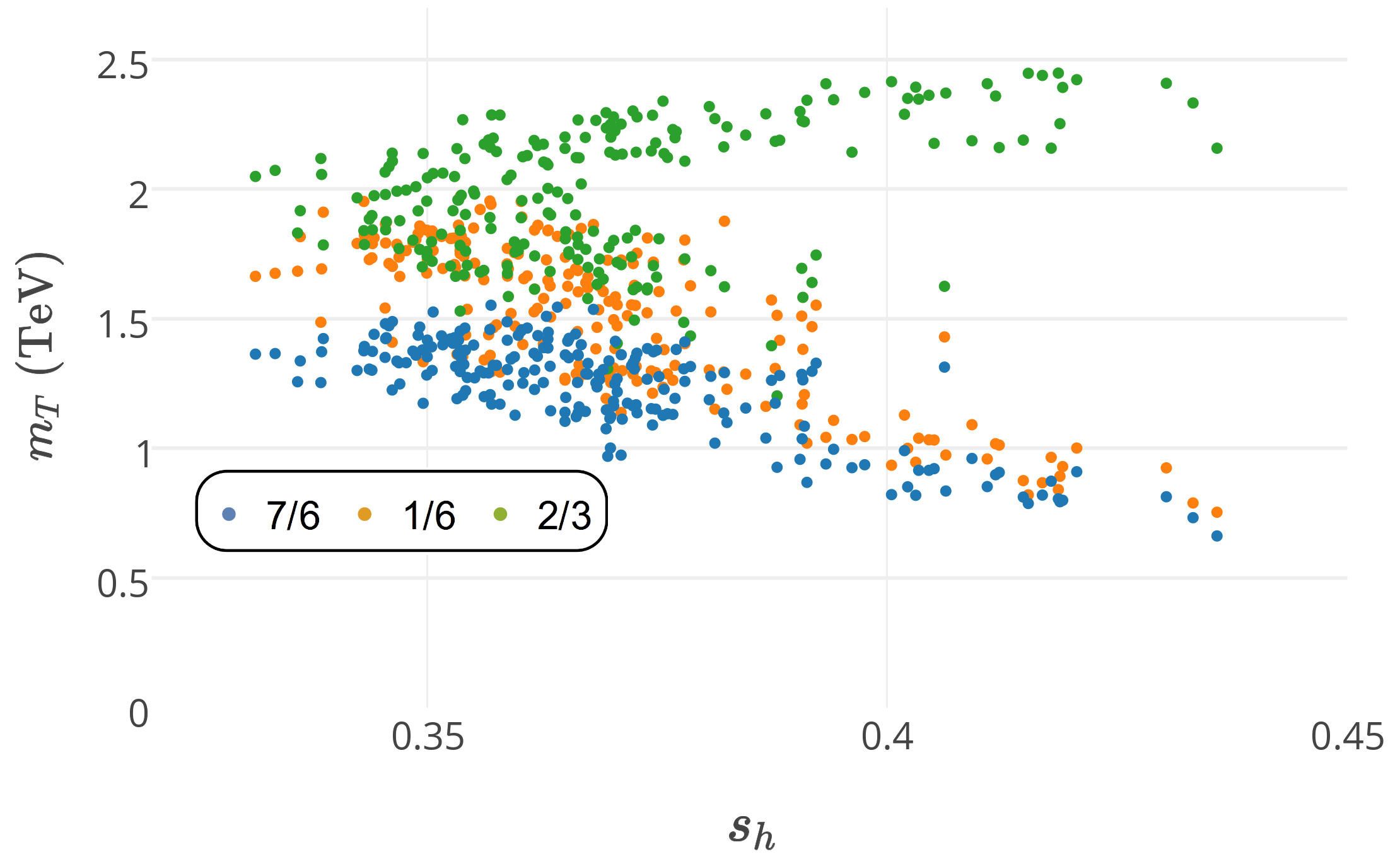}
  \caption{Correlation between $s_h$ and the top partner masses when $|m_u|<1.4$. As above, the different coloured points correspond to top partners with different hypercharge.\label{fig:shmTB}}
\end{minipage}%
\hfill
\begin{minipage}{.48\textwidth}
  \includegraphics[width=\textwidth]{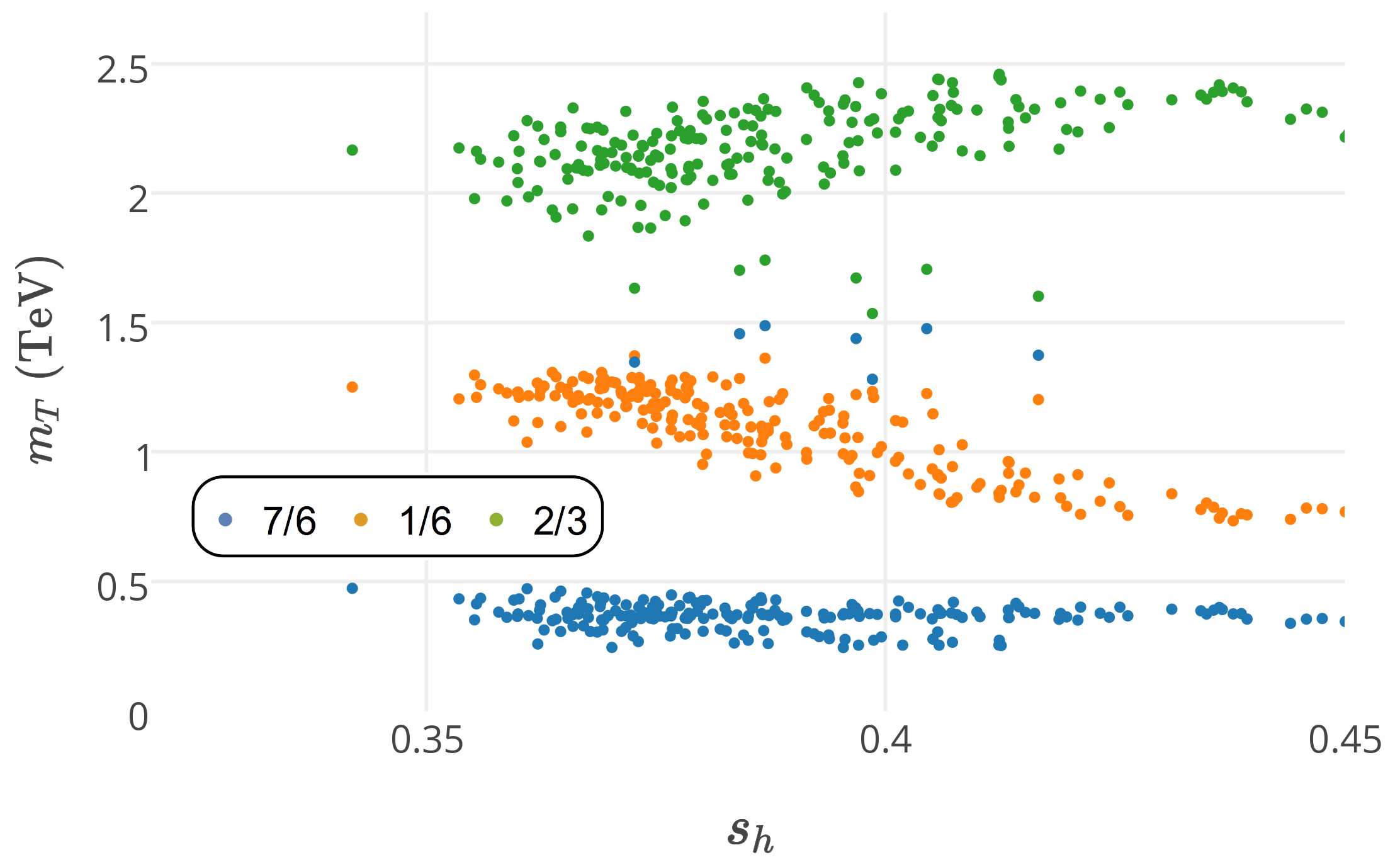}
  \caption{Correlation between $s_h$ and the top partner masses when $|m_u|>1.4$.  \label{fig:shmTA}}
\end{minipage}
\end{figure}

Since we fix $M_{KK}=1.1$ TeV and fix the vev, varying $\ln(\Omega)$ is analogous to varying $s_h$.  In figures~\ref{fig:shmTB} and~\ref{fig:shmTA} we see the correlation between top partner masses and $s_h$ explicitly.
It is clear that in figure~\ref{fig:shmTB}, reducing the mass gap between the top partners is strongly correlated with a reduction in $s_h$.  However we do not see this behaviour in figure~\ref{fig:shmTA}.  Thus from the above figures we can conclude that, when $|m_u|\lesssim1.4$ we can have less composite mixing and a smaller $s_h$ is correlated with a smaller mass gap among the top partners, and an increase in the mass of the lightest top partner.  Whereas for $|m_u|\gtrsim1.4$, we are forced to have a larger composite mixing, and lowering $s_h$ doesn't alter the top partner spectrum very much.  

\begin{figure}[H]
\centering
\begin{minipage}{.48\textwidth}
  \includegraphics[width=\textwidth]{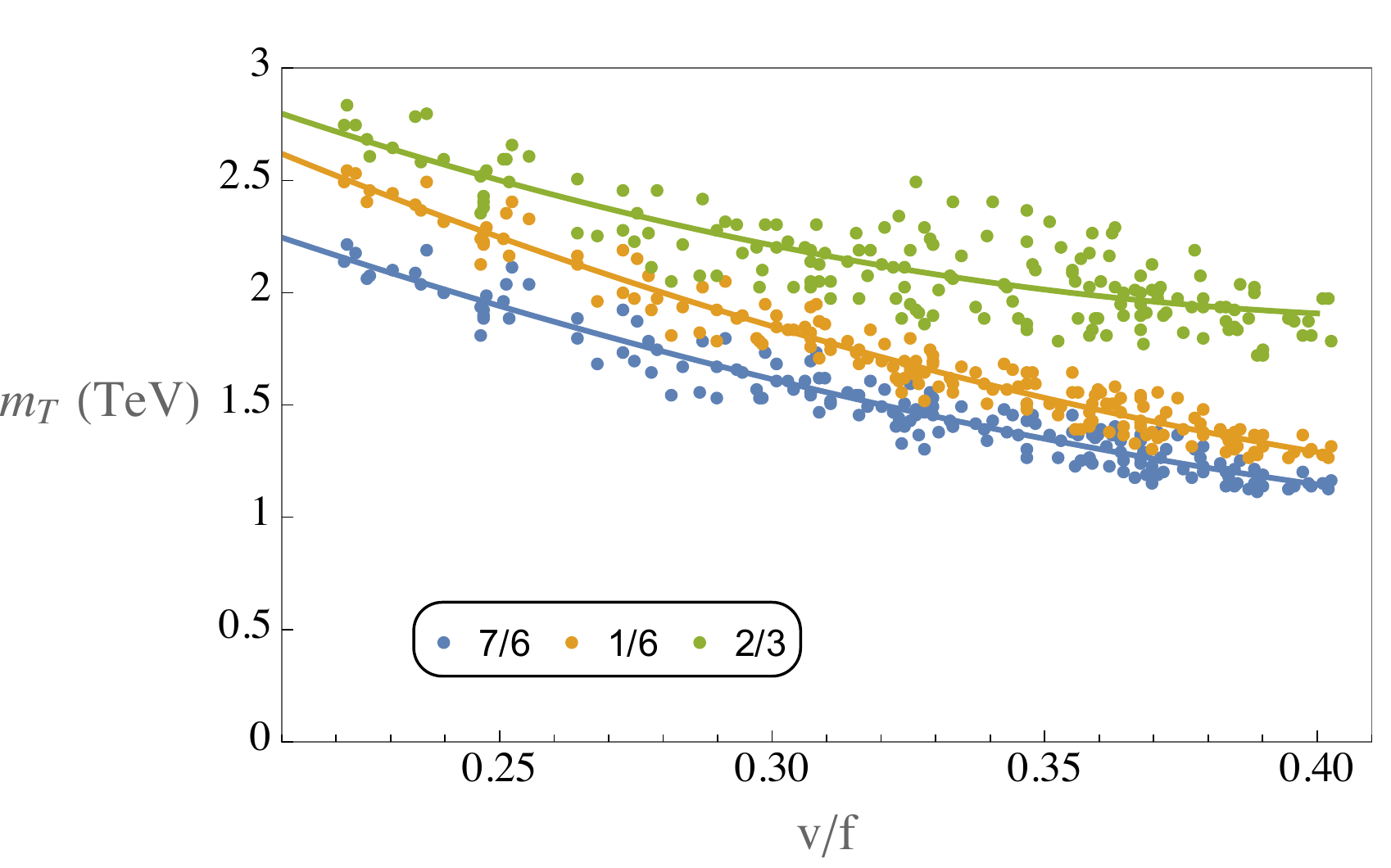}
  \caption{$c_q=0.4$, $0\leq -c_u\leq0.4$, $1\leq M_{KK}\text{(TeV)}\leq2$ TeV, $20\lesssim\ln(\Omega)\lesssim 30$ and $m_u=-1/M_u$. As above, the different coloured points correspond to top partners with different hypercharge. \label{fig:Scan3mTVF}}
\end{minipage}%
\hfill
\begin{minipage}{.48\textwidth}
  \includegraphics[width=\textwidth]{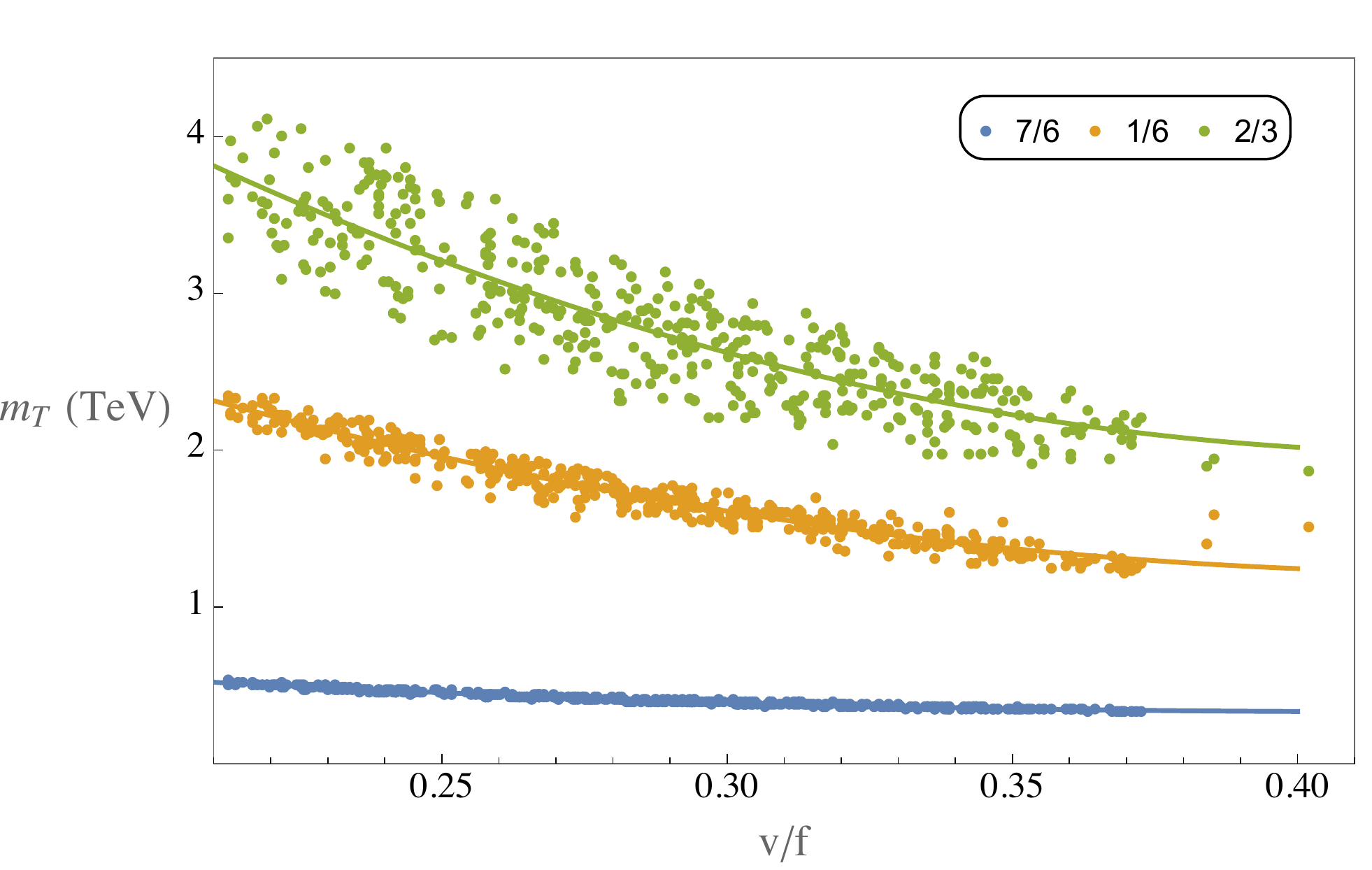}
  \caption{$c_q=0.2$, $0\leq c_u\leq0.4$, $1\leq M_{KK}\text{(TeV)}\leq2$ TeV, $20\lesssim\ln(\Omega)\lesssim 30$ and $m_u=-1/M_u$. \label{fig:Scan4mTVF}}
\end{minipage}
\end{figure}

In figures~\ref{fig:Scan3mTVF} and~\ref{fig:Scan4mTVF} we perform similar scans, except we allow $M_{KK}$ to vary.  In one case, we have a very light top partner with a large mass gap, and in the other we have no light top partners and a small mass gap.
One would naturally expect that by reducing $s_h$, the mass of the top partners increase.  What we show here is that this is only true in the case that $0\leq -c_u\leq0.4$, i.e. when there is less composite mixing for $\xi_u$.  When $0\leq c_u\leq0.4$, i.e. large composite mixing, we clearly show that lowering $s_h$ does not result in an increase in the mass of the lightest state.  This is hinted at in figure~\ref{fig:shmTA}, and re-enforced by the data in figure~\ref{fig:Scan4mTVF}.  

\begin{figure}[H]
\centering
\begin{minipage}{.48\textwidth}
  \includegraphics[width=\textwidth]{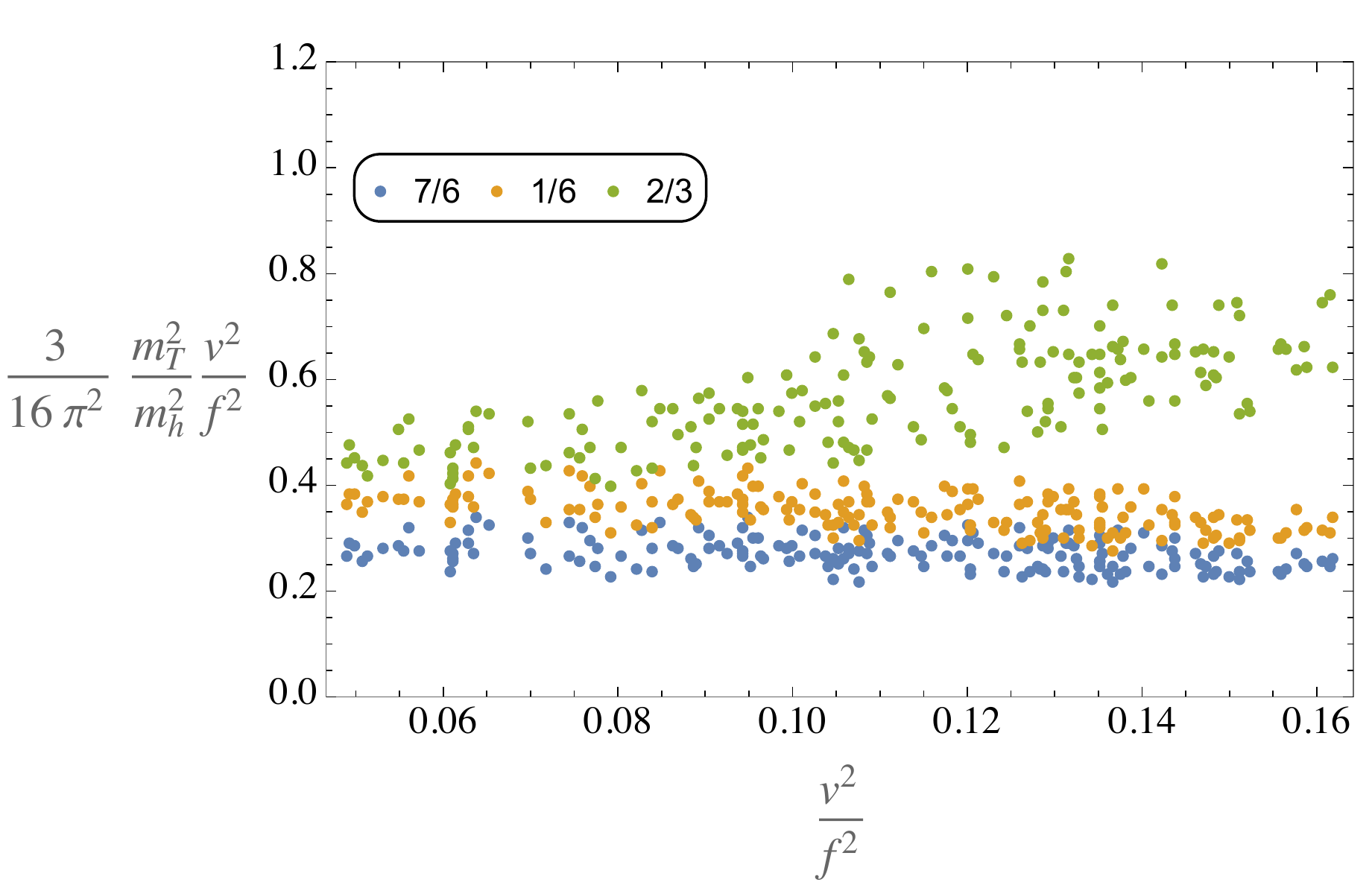}
  \caption{$c_q=0.4$, $0\leq c_u\leq0.4$, $1\leq M_{KK}\text{(TeV)}\leq2$ TeV, $20\lesssim\ln(\Omega)\lesssim 30$ and $m_u=-1/M_u$. As above, the different coloured points correspond to top partners with different hypercharge. \label{fig:Scan3mTVF2}}
\end{minipage}%
\hfill
\begin{minipage}{.48\textwidth}
  \includegraphics[width=\textwidth]{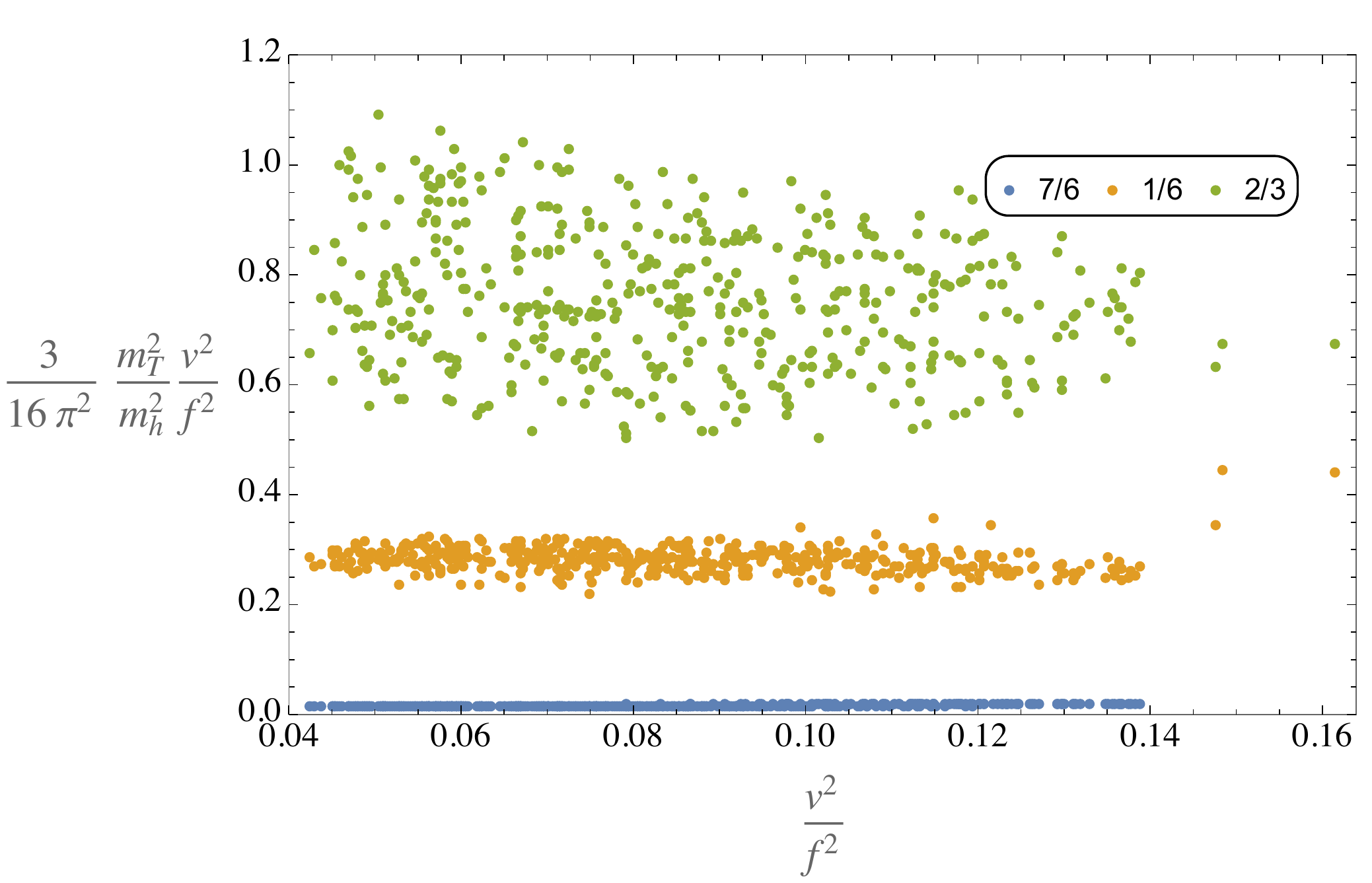}
  \caption{$c_q=0.2$, $0\leq-c_u\leq0.4$, $1\leq M_{KK}\text{(TeV)}\leq2$ TeV, $20\lesssim\ln(\Omega)\lesssim 30$ and $m_u=-1/M_u$. \label{fig:Scan4mTVF2}}
\end{minipage}
\end{figure}

Finally, let us comment on the relation between the top-partner masses and the loop contributions to the Higgs mass. 
In studying composite Higgs models in 4D it is found that one expects the following approximate relation to hold,
\begin{equation}\label{eq:4DmhmT}
m_H^2\sim \frac{3}{16\pi^2}\left(\frac{v}{f_{\pi}}\right)^2m_T^2
\end{equation} 
where $m_T$ is the mass of the top partners. Since we fix $v$ to its SM value, this implies a linear relation between the Higgs mass and both the top partner masses and the ratio $v/f$. In the  figures~\ref{fig:Scan3mTVF2} and~\ref{fig:Scan4mTVF2} test we test the latter relation, finding that this relation receives {\cal O}(1) corrections in the dual model.

\subsection{Deviations in the top Yukawa coupling}

\begin{figure}[H]\label{fig:kappavkappat}
\centering
  \includegraphics[width= 250pt]{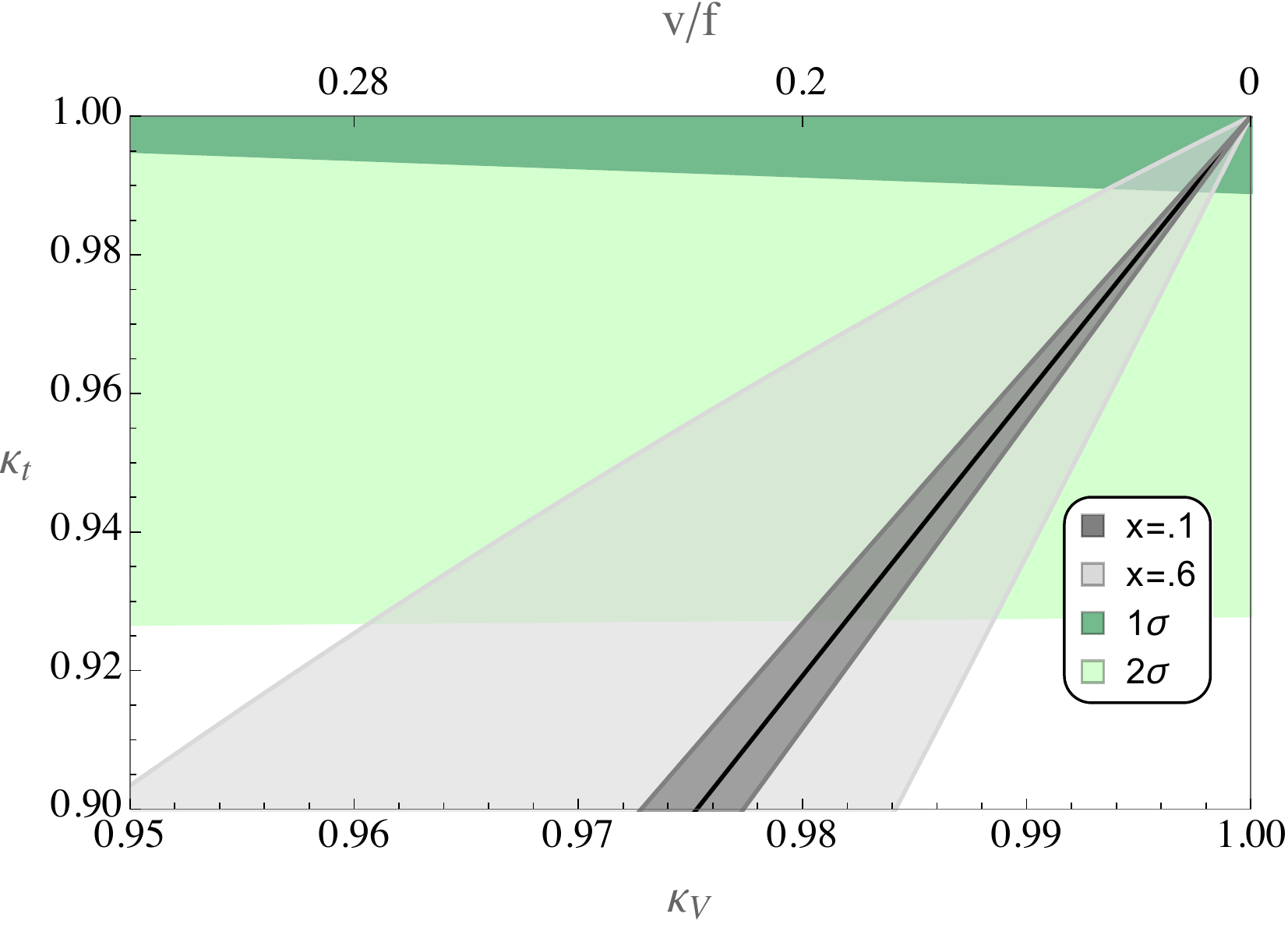}
  \caption{{Profile contribution to the Yukawa coupling: in terms of the experimental variables $\kappa_t$ and $\kappa_V$. In light and dark green the ATLAS 1$\sigma$ and 2$\sigma$ limits from \cite{Aad:2015pla}.}}\label{kappavkappat}
\end{figure}

We expect an inverse scaling between $M_u$ and (the negative of) $m_u$. We will take a mildly more general relation
\begin{equation} M_u=-\frac{a_1}{m_u} \end{equation}\label{eq:braneMass}
with $a_1$ a real constant. In this case our expression simplifies to
\begin{align}
&\frac{y_{\psi}^{(0)}v}{m_{\psi}^{(0)}}= 
1 -\frac{s_h^2}{c_h^2} - s_h^2 \left(a_1^2-1\right) \left(\frac{\tilde{c}_q}{2 a_1^2 \tilde{c}_q +2 (2-\tilde{c}_u) M_u^2}-\frac{\tilde{c}_u}{(2-\tilde{c}_q) M_u^2+\tilde{c}_u}
\right) + O\left(s_h^4 \right)
\end{align}
It is now obvious that the additional Yukawa correction due to 5D effects vanishes for either $a_1 = \pm 1$, and for flat profiles. It is also seen that the contribution switches sign for $a_1^2 =1$ and for 
$$a_1^2 =\frac{1}{2} + \frac{  M_u^2 \left((2 - \tilde{c}_q) \tilde{c}_q - 2 (2 - \tilde{c}_u) \tilde{c}_u \right) }{2 \tilde{c}_q \tilde{c}_u}$$
In other words, in the region
$$ \frac{1}{2} + \frac{  M_u^2 \left((2 - \tilde{c}_q) \tilde{c}_q - 2 (2 - \tilde{c}_u) \tilde{c}_u \right) }{2 \tilde{c}_q \tilde{c}_u} < a_1^2 < 1$$
there can be an effective cancelation between the universal contribution and the Yukawa contribution. 
\begin{figure*}[h]\label{fig:yukawasx}
    \centering
    \begin{subfigure}[t]{0.5\textwidth}
        \centering
        \includegraphics[height=2in]{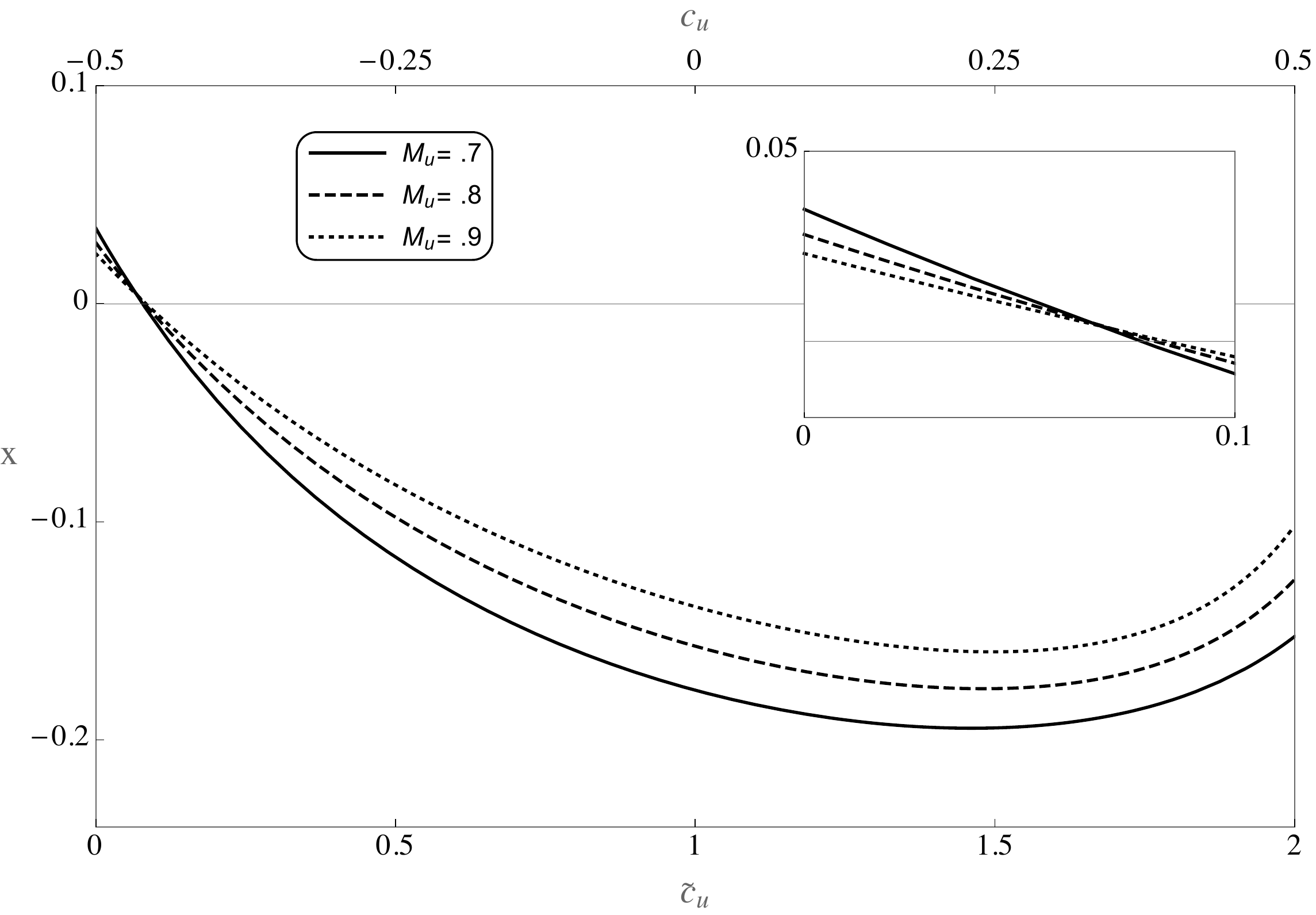}
        \caption{$c_q = 0.4$ ($\tilde{c}_q = 0.2$) and $a_1 = 1.2$}
    \end{subfigure}%
    ~ 
    \begin{subfigure}[t]{0.5\textwidth}
        \centering
        \includegraphics[height=2in]{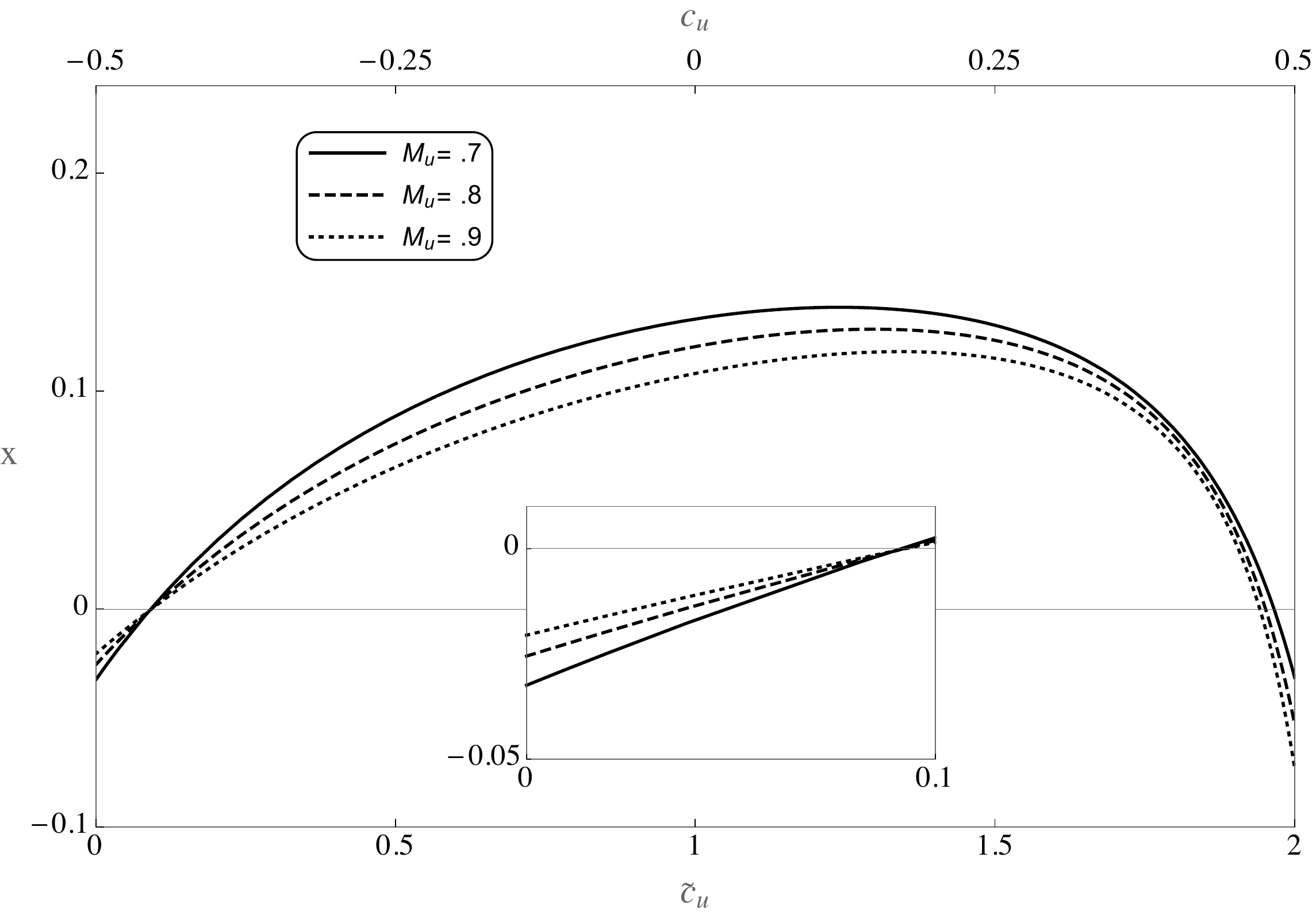}
        \caption{$c_q = 0.4$ ($\tilde{c}_q = 0.2$) and $a_1 = 0.8$}
    \end{subfigure}
    \caption{{Profile contribution to the Yukawa coupling: It is seen that the contribution is larger for IR localised fermions, and that the sign is dependent on the sign of $(a_1 - 1)$. The values of $M_u$ are chosen such that the scan results will map between the curves. }}\label{yukawa}
\end{figure*}
We can see this explicitly for two benchmark scenarios, $a_1 = 0.8$ and $a_1 = 1.2$.
Writing 
\begin{align}
&\frac{y_{\psi}^{(0)}v}{m_{\psi}^{(0)}}= 
1  - s_h^2 \left(\frac{1}{c_h^2} - x\right) + O\left(s_h^4 \right)
\end{align}
where $x$ is the Yukawa correction (modulo $s_h^{-2}$),
$$x =  \left(1- a_1^2\right) \left(\frac{\tilde{c}_q}{2 a_1^2 \tilde{c}_q +2 (2-\tilde{c}_u) M_u^2}-\frac{\tilde{c}_u}{(2-\tilde{c}_q) M_u^2+\tilde{c}_u}
\right).$$
We plot this isolated mode contribution for the benchmarks in figure~\ref{yukawa}. Here we see indeed that the sign of the correction is dependent on the sign of $a_1-1$, that is, on the relation between the brane masses $M_u$ and $m_u$. It is also seen that the correction is expected to be out of experimental reach for a small departure of $a_1 = 1$. However, the contribution can be made more sizeable values of $a_1$. For instance, in the case in which $a_1 = 1.5$, one finds a maximum of $x=0.6$ for $\tilde{c}_u \approx 1.7 $.
We use this large case to plot the range of imaginable contributions in the $\kappa_V-\kappa_t$ plane in figure~\ref{kappavkappat}. 

\subsection{Higgs differential distributions as a test of compositeness}

In equations~\ref{eq:mT} and~\ref{eq:mEWSB} we see the combinations of form factors whose zeros correspond to the top partner masses.  The expressions also contain information on the two point functions for these fields away from $p^2=m_n^2$, where $n$ denotes the $n^{th}$ resonance. In principle, one should be able to see the effect of these resonances in the form factor of the coupling of the Higgs to tops and top-partners. To produce the Higgs  with some inherent momentum, we can produce the Higgs in, e.g., association with a vector boson or  with a hard jet,
\begin{equation}
p p \to  V H \textrm{ where } V = Z \textrm{, } W^\pm \textrm{ or } p p \to H + j 
\end{equation} 
Differential distributions of, e.g., the Higgs $p_T$ would be a good proxy to understand this form factor.  In \cite{Banfi:2013yoa,Azatov:2013xha,Grojean:2013nya} the authors have studied, using 4D realisations of Composite Higgs models, the effects of the top partners in the differential distribution of the Higgs $p_T$ for the process $p p \to H + j$.  In these studies the authors only include the effects of one top-partner, with the Yukawa couplings fixed by a mixing between the top and top-partner.  This cross-section is proportional to the Yukawa couplings and is suppressed at high energies by the PDFs of the initial state gluons.  They find that the presence of top partners has a visible effect in this differential distribution, and that this technique can be used to probe a large range of top partner masses.  The method outlined there is useful for studying the effects of new heavy states on the Higgs production, but it does not include effects arising in the Higgs couplings due to the compositeness of the fields.  This can only be done if one can determine the momentum dependence of the Higgs couplings, and one advantage of the 5D holographic realisations is that they allow us to do this.  The momentum dependence is encoded in the form factors we discussed in section 3, and the effects of all top-partners are accounted for in these terms.

\begin{figure}[H]
\centering
\begin{minipage}{.75\textwidth}
  \includegraphics[width=\textwidth]{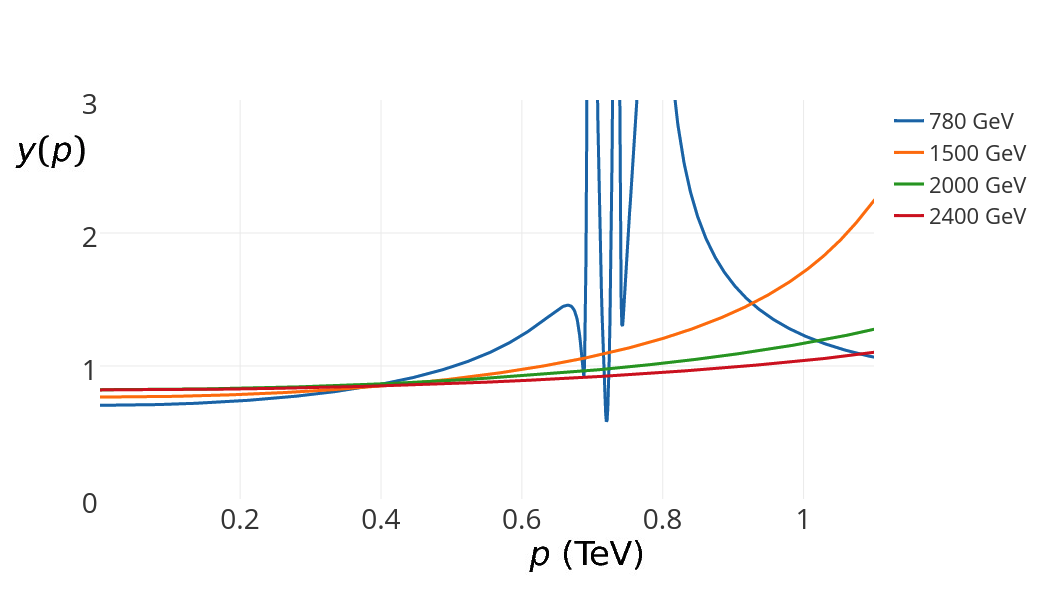}
  \caption{Here we plot the momentum dependence of the form factor for the $Ht_Lt_R$ coupling.  The masses quoted in the legend are for the hypercharge-$1/6$ top-partners, however the effects of other top-partners are also seen in the coupling.\label{fig:YukawaMomPlot}}
\end{minipage}%
\end{figure}

In figure \ref{fig:YukawaMomPlot} we plot the momentum dependence of the $Ht_Lt_R$ form factor.  We look at cases where the lightest hypercharge-$1/6$ top partner ranges from $\sim 780$ GeV to $\sim2400$ GeV, while reproducing the correct Higgs mass, top mass and v.e.v..  At low momenta we see the coupling settles at a constant value close to one, as expected.  However at larger momenta, near the top-partner masses, we see that the resonances are actually visible in the momentum dependence of this coupling.  Thus, one would imagine that this effect could be seen in the differential distribution of the Higgs $p_T$ for $gg \to H+j$.

In another work \cite{Banfi:2016inprep} we are using these form factor techniques to perform a similar analysis as done in the previous works.  The purpose of this is two-fold; firstly we will be able to include the effects of the whole tower of top-partners and the momentum dependence of the coupling in the calculation, and secondly, this will allow us to directly compare collider predictions from the 4D and 5D realisations of Composite Higgs models.

\section{Conclusions}
In this paper we addressed the question whether a light Higgs implies light top partners in the Minimal Composite Higgs model MCHM$_5$. The experimental constraints can be avoided by increasing the scale $M_{KK}$, but this is at the cost of a severe fine-tuning. Attempts at realising MCHM$_5$ without light top partners naturally have been primarily focussed on the fermion sector: 4D approaches include a different embedding of the third generation of quarks in representations of SO(5); holographic realisations include leptonic contributions to the Higgs potential. Here we propose an alternative method to alleviate the tension: we show that if the degree of composite mixing in the multiplets is reduced, the mass of the lightest top parters can be increased, without increasing the compositeness scale M$_{KK}$. To maintain a light Higgs, the cutoff in the 5D model (measured by $\ln \Omega$) is reduced. 

With an eye to the next LHC run we discuss the phenomenology of this version of the MCHM$_5$. In anticipation of improved LHC constraints on the lightest top Yukawa coupling, we show that a deviation from the relation between IR brane masses $m_u = - 1/M_u$ can reduce or enhance the Composite Higgs prediction for $y_t$ as derived from symmetry arguments alone. The deviation from the Standard Model is captured in the parameters $\kappa_V$ and $\kappa_T$, which allow for a comparison with the ATLAS data. In particular, it is seen that relaxing the brane mass relation may relieve the tension slightly by increasing the predicted coupling.

We further discussed the expected phenomenology of the top partner states in future searches. Testing the relation between the Higgs and top partner masses as a function of $v/f$, we find that the masses scale approximately linearly, as expected, with a slight deviation for the (2/3) exotic state. 

The form factors computed in the 5D dual contain qualitative information about the spectrum of top partners. In particular, in the last section we show the momentum dependence of the form factor encoding the $Ht_Lt_R$ coupling, upon which the differential distribution of Higgs $p_T$ in $p p \to H + j$ will strongly depend.  Future searches at the LHC are expected to contain decisive information about the state of the MCHM$_5$, both through measurements of the top yukawa coupling and through differential distributions of the Higgs momentum.

\appendix

\section{Form factors in the holographic $\text{MCHM}_5$}
In this appendix we present the explicit forms of the form factors introduced in section 3.3, we follow a similar procedure as in \cite{Archer:2014qga}.  Neglecting brane kinetic terms, the form factors for the gauge interactions can be written as,
\begin{equation}
\Pi^{(+)}(p)=p\frac{\mathbf{Y}_0(pR^{\prime})\mathbf{J}_0(pR)-\mathbf{J}_0(pR^{\prime})\mathbf{Y}_0(pR)}{\mathbf{Y}_0(pR^{\prime})\mathbf{J}_1(pR)-\mathbf{J}_0(pR^{\prime})\mathbf{Y}_1(pR)}
\end{equation}
\begin{equation}
\Pi^{(-)}(p)=p\frac{\mathbf{Y}_1(pR^{\prime})\mathbf{J}_0(pR)-\mathbf{J}_1(pR^{\prime})\mathbf{Y}_0(pR)}{\mathbf{Y}_1(pR^{\prime})\mathbf{J}_1(pR)-\mathbf{J}_1(pR^{\prime})\mathbf{Y}_1(pR)},
\end{equation}
and are sometimes written in terms of $\Pi_0=\Pi^{(+)}$ and $\Pi_1=(\Pi^{(-)}-\Pi^{(+)})$.

The fermionic form factors are more complicated due to the brane mixings in the IR.  We use the following holographic profiles as building blocks,
\begin{equation}
G^{+}(r,c)=\sqrt{r}\left( \mathbf{Y}_{c-\frac{1}{2}}(pR^{\prime})   \mathbf{J}_{c+\frac{1}{2}}(pr) -  \mathbf{J}_{c-\frac{1}{2}}(pR^{\prime})   \mathbf{Y}_{c+\frac{1}{2}}(pr) \right)
\end{equation}
\begin{equation}
G^{-}(r,c)=\sqrt{r}\left( \mathbf{Y}_{c-\frac{1}{2}}(pR^{\prime})   \mathbf{J}_{c-\frac{1}{2}}(pr) -  \mathbf{J}_{c-\frac{1}{2}}(pR^{\prime})   \mathbf{Y}_{c-\frac{1}{2}}(pr) \right),
\end{equation}
where $c=\pm c_{q,u}$ represents the 5D fermion mass parameter, and $q$ and $u$ represent the appropriate fermion multiplets. From now on we denote $G^{\pm}(R,c)$ simply as $G^{\pm}(c)$.  Assuming no brane kinetic terms, and only two quark multiplets with real mixings, we can write the form factors as,
\begin{equation}
\Pi^q_0(p)=\frac{1}{p}\frac{G^{+}(-c_u) G^{-}(c_q)+m_u^2G^{-}(c_u)G^{+}(-c_q)}{G^{+}(c_q)G^{+}(-c_u)-m_u^2G^{-}(-c_q)G^{-}(c_u)}
\end{equation}
\begin{equation}
\Pi^u_0(p) = -\frac{1}{p}\frac{G^{+}(c_u) G^{-}(c_q)+M_u^2G^{-}(c_u)G^{+}(c_q)}{G^{-}(c_q)G^{-}(c_u)-M_u^2G^{+}(c_q)G^{+}(-c_u)}
\end{equation}
\begin{equation}
M^{u}_0(p)=\frac{1}{2}\frac{m_u}{p}\frac{G^{+}(c_q) G^{+}(-c_q)+G^{-}(c_q) G^{-}(-c_q)+G^{+}(c_u) G^{+}(-c_u)+G^{-}(c_u) G^{-}(-c_u)}{G^{+}(c_q)G^{+}(-c_u)-m_u^2G^{-}(-c_q)G^{-}(c_u)}
\end{equation}
\begin{equation}
\Pi^{q}_1(p)=\Pi^q_0\left(m_u\rightarrow\frac{1}{M_u}\right)-\Pi^q_0
\end{equation}
\begin{equation}
\Pi^{u}_1(p)=\Pi^u_0\left(M_u\rightarrow\frac{1}{m_u}\right)-\Pi^u_0
\end{equation}
\begin{equation}
M^u_1(p)=M^u_0-M^u_0\left(m_u\rightarrow\frac{1}{M_u}\right).
\end{equation}
It is clear now that $\Pi^{q,u}_{1}\rightarrow 0$ when $m_u\rightarrow \pm \frac{1}{M_u}$ and $M^u_1\rightarrow 0$ when $m_u\rightarrow \frac{1}{M_u}$.  To get the Wick rotated form factors one simply has to rotate $p\rightarrow ip_E$, the resulting form factors are expressed in terms of modified Bessel functions $\mathbf{I}_{\alpha}$ and $\mathbf{K}_{\alpha}$.

\bibliographystyle{JHEP}
\bibliography{BibTeXReferenceList2}{}

\end{document}